\documentstyle[12pt]{article}

\font\blackboard=msbm10 at 12pt
\font\blackboards=msbm7
\font\blackboardss=msbm5
\newfam\black
\textfont\black=\blackboard
\scriptfont\black=\blackboards
\scriptscriptfont\black=\blackboardss

\newcommand{\junk}[1]{}

\newcommand{\ba}{\begin{array}}
\newcommand{\ea}{\end{array}}
\newcommand{\be}{\begin{equation}}
\newcommand{\ee}{\end{equation}}
\newcommand{\bea}{\begin{eqnarray}}
\newcommand{\eea}{\end{eqnarray}}
\newcommand{\beas}{\begin{eqnarray*}}
\newcommand{\eeas}{\end{eqnarray*}}

\def\half{{1 \over 2}}
\def\identity{{\rlap{1} \hskip 1.6pt \hbox{1}}}

\def\laplace{{\kern1pt\vbox{\hrule height 1.2pt\hbox{\vrule width
1.2pt\hskip
  3pt\vbox{\vskip 6pt}\hskip 3pt\vrule width 0.6pt}\hrule height
  0.6pt}
  \kern1pt}}
\def\scriptlap{{\kern1pt\vbox{\hrule height 0.8pt\hbox{\vrule width
  0.8pt
  \hskip2pt\vbox{\vskip 4pt}\hskip 2pt\vrule width 0.4pt}\hrule height
  0.4pt}
  \kern1pt}}
\def\slash#1{{\rlap{$#1$} \thinspace /}}
\def\roughly#1{\raise.3ex\hbox{$#1$\kern-.75em\lower1ex\hbox{$\sim$}}}

\def\str{{\rm STr} \,}
\def\sym{{\rm Sym} \,}

\def\tr{{\rm Tr} \,}

\newcommand{\NP}{{\em Nucl.\ Phys.\ }}

\newcommand{\PL}{{\em Phys.\ Lett.\ }}
\newcommand{\PR}{{\em Phys.\ Rev.\ }}

\newcommand{\MPL}{{\em Mod.\ Phys.\ Lett.\ }}
\newcommand{\PRL}{{\em Phys.\ Rev.\ Lett.\ }}

\newcommand{\gone}[1]{}
\begin{document}
\pagestyle{plain}
\setcounter{page}{1}

\baselineskip16pt

\begin{titlepage}

\begin{flushright}
PUPT-1858\\
MIT-CTP-2852\\
hep-th/9904095
\end{flushright}
\vspace{8 mm}

\begin{center}

{\Large \bf Multiple D0-branes in Weakly Curved Backgrounds\\}
%\vspace{3mm}

\end{center}

\vspace{7 mm}

\begin{center}

Washington Taylor IV$^a$ and Mark Van Raamsdonk$^b$

\vspace{3mm}
${}^a${\small \sl Center for Theoretical Physics} \\
{\small \sl MIT, Bldg. 6-306} \\
{\small \sl Cambridge, MA 02139, U.S.A.} \\
{\small \tt wati@mit.edu}\\

\vspace{3mm}
${}^b${\small \sl Department of Physics} \\
{\small \sl Joseph Henry Laboratories} \\
{\small \sl Princeton University} \\
{\small \sl Princeton, New Jersey 08544, U.S.A.} \\
{\small \tt mav@princeton.edu}
\end{center}

\vspace{8 mm}

\begin{abstract}
We investigate further our recent proposal for the form of the matrix
theory action in weak background fields.  Using Seiberg's scaling
argument we relate the matrix theory action to a low-energy system of
many D0-branes in an arbitrary but weak NS-NS and R-R background.  The
resulting multiple D0-brane action agrees with the known Born-Infeld
action in the case of a single brane and gives an explicit formulation
of many additional terms which appear in the multiple brane action.  The
linear coupling to an arbitrary background metric satisfies the
nontrivial consistency condition suggested by Douglas that the masses of
off-diagonal fields are given by the geodesic distance between the
corresponding pair of D0-branes.  This agreement arises from
combinatorial factors which depend upon the symmetrized trace ordering
prescription found earlier for higher moments of the matrix theory
stress-energy tensor.  We study the effect of a weak background metric
on two graviton interactions and find that our formalism agrees with the
results expected from supergravity.  The results presented here can be
T-dualized to give explicit formulae for the operators in any D-brane
world-volume theory which couple linearly to bulk gravitational fields
and their derivatives.

\end{abstract}

%\vspace{2cm}
\vspace{1cm}
\begin{flushleft}
March 1999
\end{flushleft}
\end{titlepage}
\newpage

\section{Introduction}

Over the past few years our understanding of string theory has
developed considerably.  We now know that the five superstring
theories as well as low-energy 11-dimensional supergravity are related
through an intricate series of dualities and it has been argued that
all these theories are limits of an underlying 11-dimensional theory
called ``M-theory'' \cite{Witten-various} whose microscopic
description is as yet unknown.  It has been found that in addition to
one-dimensional stringlike excitations there are higher-dimensional
branes in each of these theories which may in some regimes be
considered to be just as fundamental as the strings.  In the five
superstring theories there are D-branes of various dimensions
\cite{Polchinski} as well as the fundamental string and NS5-branes.
In M-theory there are M2-branes and M5-branes which are related to the
branes of the superstring theories through various duality
transformations.

A fundamental class of problems is the identification of the
world-volume action for the various branes appearing in the six
theories of interest.  This problem can be posed in a number of
contexts of differing degrees of complexity.  The simplest problem is
to find the low-energy action for a single brane in a flat background
metric with no nontrivial background supergravity fields.  A more
difficult problem is to find the action for a single brane in an
arbitrary background metric and field configuration which satisfies
the supergravity equations of motion.  The problem can be made still
harder by considering systems of many branes, either in a flat or
general background.  Even for a single fundamental superstring the
action in a general background including arbitrary R-R fields is not
yet well understood; for recent work in this direction see, for
example, \cite{bvw} and references therein.  For single D-branes the
situation is somewhat better.  The action for a single D-brane moving
in a general background is the Born-Infeld action \cite{Leigh}, which
reduces to the maximally supersymmetric $U(1)$ super Yang-Mills theory
on the world volume in the case where the brane is almost flat and has
only low-energy excitations.  This action is supplemented by
Chern-Simons type couplings to background R-R fields
\cite{Douglas,Li-bound}.  Even for the single D-brane, however, there
are subtle issues involved in giving a world-volume supersymmetric
description of the Born-Infeld theory.  For
systems of $N$  D-branes, it is known that the low-energy
action for parallel branes in a flat background is given by the
supersymmetric $U(N)$ super Yang-Mills theory found by dimensional
reduction from 10D \cite{Witten-bound}.  So far there has been little
progress in describing the action governing systems of many D-branes
in a general background.  This problem is due in part to the absence
of a nonabelian generalization of the Born-Infeld action (although one
proposal for such an action was made in \cite{Tseytlin}), and in part
to ordering problems which arise even in the low-energy theory in the
presence of general backgrounds.

In this paper we consider the simplest system of many D-branes in a
general background: low-energy configurations of many D0-branes moving
in an arbitrary but weak background of type IIA supergravity.
According to an argument of Seiberg \cite{Seiberg-DLCQ} (see also
\cite{Sen-DLCQ}), the action for such a system of
D0-branes should be related to the DLCQ description of M-theory in an
associated 11-dimensional supergravity background.  This generalizes
the BFSS matrix theory conjecture \cite{BFSS,Susskind-DLCQ}, which
states that supersymmetric matrix quantum mechanics (the low-energy
theory of $N$ D0-branes in flat space) gives a light-front description
of M-theory in a flat background.  In a previous paper
\cite{Mark-Wati-3} we used a matrix theory formulation of the
multipole moments of the supercurrent components in 11D supergravity
(derived in \cite{Mark-Wati,Dan-Wati-2,Mark-Wati-3}) to propose an
explicit description of the matrix theory action up to terms linear in
the background fields, as well as an algorithm for using higher-loop
calculations in matrix theory to find the higher order terms in the
matrix theory action in general backgrounds.  In this paper, we use
our proposal for the general background Matrix theory action and
follow the arguments of Seiberg to deduce the leading terms in the 
action
for multiple D0-branes in weak type IIA supergravity backgrounds. We 
then perform some simple tests of the
Matrix theory action and the related multiple D0-brane action.  In the
D0-brane case, we show that our prescription satisfies a constraint
originally suggested by Douglas \cite{Douglas-curved} that the masses
of off-diagonal matrix elements between a pair of separated D0-branes
agree with the minimal geodesic length between the D0-branes. This
property holds also in the Matrix theory case where the separated
objects are a pair of gravitons, and we  use it to show that the
leading order potential between a pair of gravitons in a weakly curved
Ricci-flat background is correctly reproduced by the proposed general 
background
Matrix theory action.

The paper is organized as follows.  In Section 2 we review our
proposal for the linear terms in the general background Matrix theory
action.  Then, using this action, we follow the arguments of Seiberg
to deduce leading terms in the action for multiple D0 branes in an
arbitrary weak type IIA supergravity background. In section 3 we
describe tests of the IIA and matrix theory actions.  We conclude in
section 4 with a discussion of related issues and comments on further
directions.

\section{Linear coupling to backgrounds}

In subsection \ref{sec:matrix-background} we recall the proposal made
in \cite{Mark-Wati-3} for the terms in the action of matrix theory
which are linear in the background fields.  In subsection
\ref{sec:0-backgrounds} we use the approach of Seiberg to relate this
matrix theory action to an action for multiple D0-branes in IIA
background fields.  This allows us to deduce the leading
terms in the multiple D0-brane action, which to the best of our
knowledge have not been previously described.

\subsection{Backgrounds in matrix theory}
\label{sec:matrix-background}

In \cite{Mark-Wati-3} we proposed that the linear effects of a general
matrix theory background with metric $g_{IJ} = \eta_{IJ} + h_{IJ}$ and
3-form field $A_{IJK}$ could be described by supplementing the flat
space matrix theory action
\begin{equation}
S_{{\rm flat}} = -{1 \over 2 R} \int d\tau \; \tr \biggl\lbrace
-D_{\tau} X_i D_{\tau} X_i + \frac{1}{2} [X_i,X_j] [X_i,X_j] +
\Theta_{\alpha} D_{\tau} \Theta_{\alpha} - \Theta_{\alpha}
\gamma^i_{\alpha \beta} [X_i,\Theta_{\beta}] \biggr\rbrace
\label{eq:flat-action}
\end{equation}
with
additional terms of the form
\begin{eqnarray}
S_{\rm weak}  & = & \int d \tau \;
\sum_{n = 0}^{\infty}  \sum_{i_1, \ldots, i_n}\frac{1}{n!} 
\left(
\frac{1}{2}
T^{IJ (i_1 \cdots i_n)}  \partial_{i_1} \cdots \partial_{i_n}  h_{IJ} 
(0)
+
J^{IJK (i_1 \cdots
i_n)}
 \partial_{i_1} \cdots \partial_{i_n} A_{IJK}  (0)
\right. \nonumber \\
& &\hspace{1.2in} \left.
+M^{IJKLMN (i_1
\cdots i_n)}  \partial_{i_1} \cdots \partial_{i_n}  A^D_{IJKLMN}  (0)
+ {\rm fermion \; terms} \right) \label{eq:general-background}
\end{eqnarray}
where $A^D$ is the dual 6-form field which satisfies at linear order 
\[
dA^D ={}^* (dA).
\]
In (\ref{eq:general-background}) the matrix expressions $T^{IJ (i_1
\cdots i_n)}, J^{IJK (i_1 \cdots
i_n)}, M^{IJKLMN (i_1
\cdots i_n)} $ are the matrix theory forms of the multipole moments of
the stress-energy tensor, membrane current and 5-brane current of 11D
supergravity.  Explicit forms for the parts of these moments depending
only on the 9 bosonic transverse matrices $X^i$
were given in \cite{Dan-Wati-2}, and the terms quadratic in the
fermions were given in \cite{Mark-Wati-3}, as well as some terms
quartic in the fermions.  For example,  the zeroeth moments of the
components of  the stress-energy tensor are given by
\begin{eqnarray}
T^{++} &=& {1 \over R}\str\left(\identity\right)\nonumber\\
T^{+i} &=& {1 \over R}\str\left(D_t X_i\right)\nonumber\\
T^{+-} &=& {1 \over R}\str\left({1 \over 2} D_t X_i D_t X_i + {1 
\over 4} 
F_{ij}  
F_{ij} + {1 \over 2} \Theta\gamma^i[X^i,\Theta]\right)\nonumber\\
T^{ij} &=& {1 \over R}\str\left( D_t X_i D_t X_j +  F_{ik}  F_{kj} 
- {1 
\over 4} 
\Theta\gamma^i[X_j,\Theta] - {1 \over 4} 
\Theta\gamma^j[X_i,\Theta]\right)\nonumber\\
T^{-i} &=& {1 \over R} \str\left({1 \over
2}D_t X_iD_t X_jD_t X_j + {1 \over 4} D_t X_i F_{jk} F_{jk} +
F_{ij} F_{jk} D_t X_k\right) \label{eq:stress-tensor}\\ & & - {1
\over 4R} \str\left(\Theta_\alpha
D_t X_k[X_m,\Theta_\beta]\right)\{\gamma^k\delta_{im}
+\gamma^i\delta_{mk} -2\gamma^m\delta_{ki} \}_{\alpha
\beta}\nonumber\\ & & - {1 \over 8R} \str\left(\Theta_{\alpha}
F_{kl}[X_m,\Theta_{\beta}]\right)\{ \gamma^{[iklm]} + 2 \gamma^{[lm]}
\delta_{ki} + 4\delta_{ki}\delta_{lm} \}_{\alpha \beta}\nonumber\\ & &
+ {i \over 8R} \tr(\Theta \gamma^{[ki]} \Theta \; \Theta \gamma^k
\Theta)\nonumber\\
T_f^{--} &=& {1 \over 4R} \str\left(F_{ab}F_{bc}F_{cd}F_{da} - {1 \over 
4}F_{ab} 
F_{ab} F_{cd} F_{cd}  + {\Theta} \Gamma^b \Gamma^c \Gamma^d F_{ab} 
F_{cd} 
D_a\Theta + {\cal O} ({\Theta^4})\right)\nonumber
\end{eqnarray}
where ${\rm STr}$ indicates a trace which is symmetrized over all
orderings of terms of the forms $F_{ab}, \Theta$ and $[X^i, \Theta]$,
indices $i (a)$ run from 1 (0) through 9, and we have defined
$F_{0i} = D_t{X}^i,
F_{ij} = i[X^i, X^j]$.  There are two types of terms which contribute
to higher moments of these components of the stress-energy tensor
\begin{equation}
T^{IJ (i_1 i_2 \cdots i_n)} = 
\sym (T^{IJ}; X^{i_1}, X^{i_2}, \ldots, X^{i_n}) + 
T_{\rm fermion}^{IJ(i_1 i_2 \cdots i_n)} 
% \label{eq:}
\end{equation}
The contributions $\sym ({\rm STr}\; (Y); X^{i_1}, \ldots, X^{i_n})$
are defined as the symmetrized average over all possible orderings
when the matrices $X^{i_k}$ are inserted into the trace of any product
$Y$ of the forms $F_{ab}, \Theta, [X^i, \Theta]$.  In general there
are additional fermionic contributions of arbitrary order to the
higher multipole moments, of which the simplest example is the spin
contribution to the angular momentum
\begin{equation}
T_{\rm  fermion}^{+i(j)} = {1 \over 8R} \tr(\Theta \gamma^{[ij]} 
\Theta)
% \label{eq:}
\end{equation}
The precise form of these fermionic contributions will not be
important to us in this paper, for reasons which will be discussed in
section \ref{sec:oscillator-masses}.

The results of \cite{Mark-Wati-3} for the matrix membrane and 5-brane
currents are reproduced in the Appendix for convenience.  With these
definitions, (\ref{eq:general-background}) gives a formulation of
matrix theory in a weak background metric to first order in the metric
$h_{IJ}$, the 3-form $A_{IJK}$, and all their higher derivatives.  It
was argued in \cite{Mark-Wati-3} that if the matrix theory conjecture
is true in flat space, this formulation must be correct at least to
order $\partial^4h, \partial^4A$ for a class of backgrounds which can
be produced as the long range fields around supergravity sources
described by matrix theory objects.  We conjectured further that this
form may work to all orders and in a general background.  It should be
emphasized, however, that this formulation can only be given for
M-theory backgrounds with a global $U(1)$ symmetry around a compact
direction, as we do not know how to incorporate dependence of the
background on the compact coordinate $x^-$.  We only expect this
action to be part of a consistent all-orders matrix theory action in a
general background when the background satisfies the equations of
motion.  The derivation of this action also depended upon a particular
choice of gauge for the graviton, so that it may be necessary to
restrict attention to background fields satisfying the linearized
gauge

\begin{equation}
\partial^I \bar{h}_{IJ} =\partial^I (h_{IJ}-\frac{1}{2} 
\eta_{IJ}h_{K}^{\; K}) = 0.
% \label{eq:}
\end{equation}

\subsection{Backgrounds for D0-branes} 
\label{sec:0-backgrounds}

We now investigate how the Matrix theory action described in the 
previous
section is related to the action for multiple D0-branes in background
type IIA supergravity fields.  

In the case of a flat background, the
Matrix theory action may be derived by showing an equivalence between
the DLCQ limit of M-theory in a flat background with N units of momentum
around the circle and a particular limit of type IIA string theory with
N D0-branes \cite{Seiberg-DLCQ}.  In this limit, the only remaining
degrees of freedom are the lowest energy modes of open strings ending on
the N D0-branes.  The dynamics of these modes are in general described
by a non-abelian generalization of the Born-Infeld action whose complete
form is not known.  However, in the appropriate limit of type IIA string
theory, most of the terms in this action vanish, and we find that the
dynamics of DLCQ M-theory in a flat background are described by an
action equivalent to the dimensional reduction of D=10 super Yang-Mills
theory  to 0+1 dimensions.

   The action for Matrix theory with background supergravity fields
given in the previous section has been derived completely within the
context of Matrix theory.  However, in principle, one should be able to
apply Seiberg's arguments to this case also and derive the same action
as a limit of the action for D0-branes in type IIA string theory with
background supergravity fields.  Again, only particular terms in the
D-brane action will survive in the appropriate limit, but unlike the
flat space case, not even these terms are known except in the case of a
single brane.  Hence, in the case $N=1$, we should be able to rederive
our result from previously known facts about D-branes, but more
importantly, we will be able to apply the arguments in reverse for $N>1$
to derive previously unknown leading terms in the action for multiple
D0-branes in an arbitrary weak type IIA supergravity background.  Using
T-duality, our result may be extended to give leading terms in the
actions for all other types of D-branes.

\subsubsection{Relationship between DLCQ and type IIA backgrounds}

   We now review the steps taken in \cite{Seiberg-DLCQ} as they apply in 
the case of weak backgrounds to make precise the relationship between 
the matrix theory action and the multiple D0-brane actions. In 
particular, we must determine the relationship between the D=11  
supergravity fields appearing in the Matrix theory action
(\ref{eq:general-background})
and the 
type IIA supergravity fields appearing in the related
D0-brane action.  

We start by considering M-theory with background metric
\[
{g}_{IJ} = \eta_{IJ} + {h}_{IJ}
\] 
in a frame with a compact coordinate $x^-$ of size $R$ which is
light-like in the flat space limit $h_{IJ} = 0$.  This theory can be
described as a limit of a family of space-like compactified theories.
Defining an $\hat{M}$-theory with background metric
\[
\hat{g}_{IJ} = \eta_{IJ} + \hat{h}_{IJ}
\] 
in a frame with a spacelike compact direction $x^{10}$ of size $R_s$,
the DLCQ limit of the original M-theory can be found by boosting the
$\hat{M}$-theory in the $x^{10}$ direction with boost parameter
\[
\gamma = \sqrt{{R^2 \over 2 R_s^2} + 1}
\]
and then taking a limit $R_s \rightarrow 0$. 
The metric $\hat{g}_{IJ}$ in the $\hat{M}$-theory
is related to that of the original M-theory by
\beas
\hat{h}_{ij} &=& h_{ij}\\
\hat{h}_{0\,i} &=& {\alpha \over \sqrt{2}}h_{+i} + {1 \over 
\alpha\sqrt{2}} h_{-i}\\
\hat{h}_{10\,i} &=& {\alpha \over \sqrt{2}}h_{+i} - {1 \over 
\alpha\sqrt{2}} h_{-i}\\ 
\hat{h}_{0\,0} &=& h_{+-} + {\alpha^2 \over 2}h_{++} + {1 \over 
2 \alpha^2}h_{--}\\ 
\hat{h}_{10 \, 10} &=& -h_{+-} + {\alpha^2 \over 2}h_{++} + {1 \over 
2 \alpha^2}h_{--}\\      
\hat{h}_{0\,10} &=&  {\alpha^2 \over 2}h_{++} - {1 \over 
2 \alpha^2}h_{--}\\
\eeas
where we have defined
\beas
\alpha &=& \gamma(1-v) = \gamma - \sqrt{\gamma^2 - 1}\\
&=& {R_s \over R \sqrt{2}} + {\cal O}((R_s/R)^3)
\eeas

M-theory on a small spacelike circle of radius $R_s$ 
is equivalent to type IIA string theory with background fields given to  
leading order by:
\beas
h^{IIA}_{\mu \nu} &=& \hat{h}_{\mu \nu} + {1 \over 2} \eta_{\mu \nu} 
\hat{h}_{10\,10}\\
C_{\mu} &=& \hat{h}_{10 \, \mu}\\
\phi &=& {3 \over 4} \hat{h}_{10 \, 10}
\eeas    
and parameters 
\[
g_s = (R_s M_p)^{3/2}, \; \; \; M_s = R_s^{1/2}M_p^{3/2}
\]
where $M_p$ is the eleven-dimensional Planck mass. Here we have defined 
$g_s$ to be a constant and chosen the dilaton $\phi$ so that $\phi=0$ in 
the case of a circle of constant size $R_s$ with $h_{10 \; 10} = 0$. 
Thus the effective string coupling is given locally by the combination
\[
g_s(\vec{x}) = g_s e^{\phi} .
\]
\\
Combining the two equivalences, we conclude that DLCQ M-theory with N 
units of momentum on a lightlike circle of size $R$ and background 
metric $g_{IJ}$ is equivalent to the $R_s \rightarrow 0$ limit of type 
IIA string theory with N D0-branes, parameters
\[
g_s = (R_s M_p)^{3/2}, \; \; \; M_s = R_s^{1/2}M_p^{3/2}
\]
and background fields
\bea
h^{IIA}_{0 0} &=& {3 \over 2}h_{+-} + {\alpha^2 \over 4}h_{++} + {1 
\over 
4 \alpha^2} h_{--} \nonumber\\
h^{IIA}_{0 i} &=&  {\alpha \over \sqrt{2}} h_{+i} + {1 \over \alpha 
\sqrt{2}} h_{-i}
\nonumber\\
h^{IIA}_{i j} &=& h_{ij} + {1 \over 2} \delta_{ij} (-h_{+-} + {\alpha^2 
\over 2}h_{++} + {1 \over 2 \alpha^2} h_{--}) \label{eq:relations}\\
\phi &=& -{3 \over 4}h_{+-} + {3 \alpha^2 \over 8}h_{++} + {3 \over 
8 \alpha^2} h_{--}\nonumber\\
C_0 &=& {\alpha^2 \over 2}h_{++} - {1 \over 2 \alpha^2} h_{--} 
\nonumber\\
C_i &=& {\alpha \over \sqrt{2}} h_{+i} - {1 \over \alpha \sqrt{2}} 
h_{-i} 
\nonumber
\eea

At first glance, such a limit seems problematic.  In particular, it
appears that for fixed finite values of the DLCQ metric components,
the background fields of the equivalent type IIA theory diverge in the
limit $R_s \rightarrow 0$ since $1/\alpha \rightarrow \infty$.  However, 
recall from the flat space case
that without a further rescaling of the parameters in the type IIA
picture, the energies of the states we are interested in go to 0 like
$R_s$.  As we shall see, the appropriate rescaling of parameters which
makes the energies we are interested in finite without changing the
physics also ensures that the apparent divergences of background field
components do not lead to divergent terms in the final action.

Another feature of this action is that after the appropriate
rescaling the characteristic length scale  $L$ associated with the
structure of the metric becomes much smaller than the string length
$1/M_s$.  While this may seem unusual, it is precisely what is needed
for the physics of the system to be completely captured by the open
string theory describing the D0-brane theory at substring scales
studied in \cite{DKPS}.  Indeed, for compact manifolds such as tori,
it is this effect which makes it possible for the wrapped string modes
corresponding to momentum excitations on the dual space to
become physically relevant \cite{WT-compact,Seiberg-DLCQ}.
   
\subsubsection{$N=1$ actions}

We now use the correspondence just discussed to make an explicit
comparison between the matrix theory and IIA descriptions of a system
of $N$ 0-branes in a weak background field.  We begin with the case
$N=1$.  Here, both the Matrix theory and D0-brane actions are known,
so we would like to check that the Matrix theory action may be derived
from the D0-brane action before proceeding to the case $N>1$ where the
D0-brane action is not known. For the case $N=1$, the Matrix theory
action (\ref{eq:flat-action},\ref{eq:general-background}) reduces to
\begin{eqnarray}
S & = & {1 \over R} \int dt \left\{ \frac{\dot{x}^2}{2} + \frac{1}{2} 
h_{++}(\vec{x}) + 
h_{+i}(\vec{x}) \dot{x}^i
+ \half h_{ij}(\vec{x}) \dot{x}^i \dot{x}^j 
\label{graviton} \right.\\
 &  &\hspace{1in} + \left.\frac{1}{2} h_{+-}(\vec{x})\dot{x}^2 + 
\frac{1}{2} 
h_{-i}(\vec{x}) \dot{x}^2 \dot{x}^i + \frac{1}{8} h_{--}(\vec{x}) 
\dot{x}^4 \right\}. \nonumber
\end{eqnarray}
In this case we expect the action to describe a single graviton in
curved space with unit momentum along the lightlike circle.  Such an
action was derived from supergravity in \cite{Mark-Wati}; expression
(\ref{graviton}) is indeed identical to the supergravity result given by
equation (13) in that paper.

The world-volume action for a single D0-brane moving in a general 
type IIA background supergravity fields is given by
\begin{equation}
S_{IIA} = - \tau_0\int d \xi e^{-\phi} \sqrt{g_{\mu \nu}
(d{x}^{\mu}/d \xi )
(d{x}^{\nu}/d \xi)}  +
\tau_0 \int C_{\mu} dx^{\mu} 
\label{eq:single-0}
\end{equation}
where $\phi$, $g_{\mu \nu}$, and $C_{\mu}$ are the background dilaton,
metric, and R-R one-form fields, and the parameter $\tau_0$
is the D0 mass, given by
\[
\tau_0 = {M_s \over g_s} .
\]
One can also consider background R-R three form $C_{\mu \nu \lambda}$ 
and 
NS-NS antisymmetric tensor $B_{\mu \nu}$ fields,
but these do not couple to a 
single zero-brane.  
   
   According to the equivalence presented in the previous section, the
Matrix theory action (\ref{graviton}) should arise from the D0-brane
action (\ref{eq:single-0}) by rewriting the type IIA background fields
in terms of the desired DLCQ supergravity background using the
relations (\ref{eq:relations}), then rescaling parameters and taking
the limit $R_s \rightarrow 0$. We will now verify this explicitly.
Choosing a gauge in which the coordinate time $x^0$ is identified with
the worldvolume time $\xi$ we first expand the D0-brane action to
leading order in the background fields, giving
\begin{equation}
S = -\tau_0 \int d\xi \left\{ (1-v^2)^{1/2} (1- \phi) - {1 \over 2} 
(1-v^2)^{-1/2}(h^{IIA}_{00} + 2h_{0i}v^i + h_{ij}v^iv^j) - C_0 - C_iv^i 
\right\}
\label{eq:expand-Born-Infeld}
\end{equation}
where $v^i \equiv \dot{x}^i$. We now write the IIA background fields
in terms of the background fields in the equivalent DLCQ M-theory
using (\ref{eq:relations}).  Keeping only the leading term in $R_s/R$
for each of the components of the metric $h_{IJ}$, we find
\begin{eqnarray}
S & = &  {1 \over R_s} \int d \xi \left\{ -1  + {1 \over 2} 
\frac{R^2_s}{R^2} 
h_{++}(\vec{x}) + \frac{R_s}{ R} 
h_{+i}(\vec{x}) v^i\right. \label{eq:replaced}\\
& &\hspace{1in} \left.
+ \half h_{ij}(\vec{x}) v^i v^j + \frac{1}{2} 
h_{+-}(\vec{x})v^2 +  \frac{v^2}{2}+\frac{1}{2} 
\frac{R}{R_s} h_{-i}(\vec{x}) v^2 v^i + {1 \over 8}
\frac{R^2}{R_s^2} h_{--}(\vec{x}) v^4 \right\}. \nonumber
\end{eqnarray}
Many of these terms seem to diverge in
the $R_s \rightarrow 0$ limit we are interested in. However, as
mentioned above, this scaling is deceptive, since we must rescale
parameters in the theory so that the energies of the states we are
interested in remain finite rather than going to zero in the limit.
Indeed, from the fact that the conjugate momentum has a leading term
of order $v/R_s$ it can be seen that all the terms in
(\ref{eq:replaced}) which are linear in the background contribute to
the Hamiltonian at order $R_s$.  Thus, as we need for the Seiberg
limit, the energy of the states of interest scale as $R_s$.

We may now perform the rescaling of \cite{Seiberg-DLCQ} by replacing
\[
R \rightarrow ({R_s \over R})^{1/2}R , \;\;\;\;\; \, \, \vec{x} 
\rightarrow ({R_s \over R})^{1/2}\vec{x}, \;\;\;\;\; \, \, h(\vec{x}) 
\rightarrow h(\vec{x}).
\]
Note that the change of variables in the second replacement combines 
with 
the rescaling of dimensionful coefficients in the expansion of $h$ to 
leave $h(\vec{x})$ unchanged, as suggested by the final replacement. 

With these redefinitions the action (\ref{eq:replaced}) becomes
\begin{eqnarray}
S & = &  \int d \xi \left\{ -\frac{1}{R_s}   +  {1 \over R}\left(
{1 \over 2} h_{++}(\vec{x}) + 
h_{+i}(\vec{x}) v^i+ \half h_{ij}(\vec{x}) v^i v^j\right. 
\right.\nonumber\\
& &\left.\left.\hspace{1.5in}
 + \frac{1}{2} 
h_{+-}(\vec{x})v^2 +  \frac{v^2}{2} +\frac{1}{2} 
h_{-i}(\vec{x}) v^2 v^i + 
 {1 \over 8} h_{--}(\vec{x}) v^4 \right)\right\}. \nonumber
\end{eqnarray}
The first  term is divergent in the $R_s \rightarrow 0$ limit and arises 
from the
BPS energy of the single 0-brane; this term also appears in the flat
space theory and is discounted in the matrix theory limit.
Dropping this term gives precisely the matrix theory action
described by (\ref{graviton}) in the $N = 1$ case.  Thus, we have
shown that the known Born-Infeld action for a single D0-brane
correctly reproduces the matrix theory action in a weak background
when the proper limit is taken.

\subsubsection{$N>1$ actions}

   We now turn to the case $N>1$.  Here, the appropriate action for
multiple D0-branes is not known, but by requiring that it reproduces the
general background Matrix theory action in the Seiberg limit, we will be
able to deduce its leading terms.

We first write down the D0-brane action in terms of the unknown
quantities coupling to the background fields.  We define quantities
$I_x$ coupling linearly to each of the background fields,
so that to leading order in the background fields, the action for $N$ D0
branes is
\bea
S  &= &  S_{{\rm flat}}  +
 \int dt  \sum_{n=0}^{\infty} {1 
\over n!} \left[
\frac{1}{2}
(\partial_{k_1}\cdots 
\partial_{k_n} h^{IIA}_{\mu \nu}) \; I_h^{\mu \nu (k_1 \cdots k_n)}
+ (\partial_{k_1}\cdots \partial_{k_n} \phi) 
\; I_{\phi}^{(k_1 \cdots k_n)}\right.\label{eq:IIA-general}\\ 
& & \hspace{1in}+ (\partial_{k_1}\cdots 
\partial_{k_n} C_{\mu }) \; I_0^{\mu (k_1 \cdots k_n)}
+
%\frac{1}{7!}
 (\partial_{k_1}\cdots 
\partial_{k_n} \tilde{C}_{\mu \nu \lambda \rho \sigma \tau \zeta }) 
\; I_6^{\mu \nu \lambda
\rho \sigma \tau \zeta (k_1 \cdots k_n)}
\nonumber\\
& & \hspace{1in}+
%\frac{1}{2}
 (\partial_{k_1}\cdots \partial_{k_n} 
B_{\mu \nu}) \; I_s^{\mu \nu (k_1 \cdots k_n)} 
+ 
%\frac{1}{6!}
(\partial_{k_1}\cdots \partial_{k_n} 
\tilde{B}_{\mu \nu \lambda \rho \sigma \tau}) \; I_5^{\mu \nu \lambda
\rho \sigma \tau
 (k_1 \cdots k_n)} \nonumber\\
& &\hspace{1in} \left.+ 
%\frac{1}{6}
(\partial_{k_1}\cdots 
\partial_{k_n} C^{(3)}_{\mu \nu \lambda }) \; I_2^{\mu \nu \lambda (k_1 
\cdots 
k_n)}
+
%\frac{1}{5!}
 (\partial_{k_1}\cdots 
\partial_{k_n} \tilde{C}^{(3)}_{\mu \nu \lambda \rho \sigma }) 
\; I_4^{\mu \nu \lambda \rho \sigma (k_1 \cdots 
k_n)}\right] \nonumber
\eea
Here, $S_{{\rm flat}}$ is the flat space action for N D0-branes, 
whose leading
terms are the dimensional reduction of D=10 SYM theory to 0+1
dimensions.  The complete form of the higher order terms in the flat
space action is not known,
but these terms vanish in the Matrix theory limit.  
We assume that the background satisfies the source-free IIA
supergravity equations of motion so that the dual fields $\tilde{C},
\tilde{B}, \tilde{C}^{(3)}$ are well-defined 7-, 6- and 5-form fields
given (at linear order) by
\begin{equation}
d \tilde{C} ={}^* dC, \;\;\;\;\;
d \tilde{B} ={}^* dB, \;\;\;\;\;
d \tilde{C}^{(3)} ={}^* dC^{(3)}.
% \label{eq:}
\end{equation}
The sources $I_{2n}$ are associated with Dirichlet $2n-$brane
currents, while the sources $I_{s}$ and $I_{5}$ are associated with
fundamental string and NS5-brane currents respectively.  It may seem
surprising that a system of D0-branes can give rise to a nonzero
D2-brane, D4-brane or D6-brane charge.  Indeed, the integrated higher 
brane
charges must vanish for a system containing a finite number $N$ of
D0-branes.  Even for $N = 2$, however, a D0-brane configuration can
have nonvanishing multipole moments of higher D-brane charges.  This
essentially arises as the T-dual of the result that the $n$th power of 
the
curvature form $F$ on a Dirichlet D$p$-brane carries $(p-2n)$-brane
charge \cite{Witten-small,Douglas}; see \cite{WT-Trieste} and
references therein for a further discussion of this issue.

The problem we address in this subsection is the determination of the
IIA currents $I_x$  under the assumption that this action 
reproduces the matrix theory action (\ref{eq:general-background}) in
the Seiberg limit.  As we will see, the leading terms in all the
currents other than $I_5$ can be determined and are related to the
matrix theory supercurrent components tabulated in the Appendix.

For the case $N=1$
we see from  (\ref{eq:expand-Born-Infeld}) that the nonvanishing
source components $I_x$ are
\bea
I_h^{00(k_1\cdots k_n)} &=& {1 \over R_s}(1-\dot{x}^2)^{-1/2} x^{k_1} 
\cdots 
x^{k_n} \nonumber\\
I_h^{0i(k_1\cdots k_n)} &=& {1 \over R_s}(1-\dot{x}^2)^{-1/2} \dot{x}^i 
x^{k_1} 
\cdots x^{k_n} \nonumber\\
I_h^{ij(k_1\cdots k_n)} &=& {1 \over R_s}(1-\dot{x}^2)^{-1/2} \dot{x}^i 
\dot{x}^j 
x^{k_1} \cdots x^{k_n} \label{eq:n1}\\
I_\phi^{(k_1\cdots k_n)} &=& {1 \over R_s}(1-\dot{x}^2)^{1/2} x^{k_1} 
\cdots 
x^{k_n} 
\nonumber\\
I_0^{0(k_1\cdots k_n)} &=& {1 \over R_s} x^{k_1} 
\cdots 
x^{k_n} \nonumber\\
I_0^{i(k_1\cdots k_n)} &=& {1 \over R_s}\dot{x}^i x^{k_1} 
\cdots 
x^{k_n} \nonumber
\eea
In the nonabelian case $N > 1$, the quantities $I_x$ will be some 
complicated 
functions of the $N \times N$ hermitian matrices $X^i$ as well as the 
fermionic matrices $\Theta$. For each $I$, we can make an expansion 
analogous to expanding in velocities for the  $N=1$ case. We write
\[
I_x = \sum I_{x[n]}
\]
where $n$ counts the dimension of a function of the matrices $X,
\Theta$, giving $X$ dimension 1, $\dot{X}$ dimension 2 and $\Theta$
dimension 3/2.  If we do a similar expansion for the flat
space action $S_0$, we find that it is precisely the $n=4$ terms
that remain in the Matrix theory limit, the higher order terms being
scaled to zero.  In the general background case, the Matrix theory
action will arise from terms $I_n$ with $n \le 8$ (and their higher
moments), so it is these
terms that our analysis will determine.

We now proceed just as in the $N=1$ case. 
We begin by working through the details of the analysis for those terms
coupling to the IIA graviton, dilaton and R-R 1-form field.  These
terms are the most complicated.  The analysis for the remaining bosonic 
fields
is described at the end of this section.

By the Seiberg equivalences, the Matrix theory action with background
supergravity fields should result from replacing the IIA background
fields in (\ref{eq:IIA-general}) with their DLCQ counterparts
(\ref{eq:relations}), rescaling parameters as above, and taking the
limit $R_s \rightarrow 0$. Before rescaling, we find that the D0-brane
action becomes
\bea
S &= & S_{\rm flat} - \frac{1}{2}\int dt \left[ (\sqrt{2} \alpha)^{-2} 
h_{--} 
\{ {1 \over 2} I_h^{00} + {1 \over 2} I_h^{ii} + {3 \over 2} I_\phi - 
2 I_0^0 \}   \right.
\nonumber\\
& &\hspace{1in} + (\sqrt{2} \alpha)^{-1} h_{-i} \{2 I_h^{0i}  - 2 I_0^i 
\}\nonumber\\
& &\hspace{1in} + h_{ij} \{ I_h^{ij} \}\nonumber\\
& &\hspace{1in} + h_{+-} \{ {3 \over 2} I_h^{00} - {1 \over 2}
I_h^{ii}  -
{3 \over 2} 
I_\phi 
\}\label{eq:IIA-rewritten}\\
& &\hspace{1in} + (\sqrt{2} \alpha) h_{+i} \{  I_h^{0i} +  I_0^i 
\}\nonumber\\
& &\hspace{1in} + (\sqrt{2} \alpha)^{2} h_{++} \{ {1 \over 8} I_h^{00} + 
{1 
\over 
8} 
I_h^{ii} + {3 \over  8} I_\phi + {1 \over  2} I_0^0 \}  \nonumber\\
& &\hspace{1in} + \left. \{ {\rm higher \; moment \;  terms} \}\right]
\nonumber
\eea
The higher moment terms have exactly the same form as the terms 
written, for example the full set of terms linear in $h_{-i}$ would be
\[
\sum_{n=0}^{\infty} {R \over R_s} \partial_{k_1} \cdots \partial_{k_n} 
h_{-i} \{2 I_h^{0i(k_1 \cdots k_n)}  - 2I_0^{i(k_1 \cdots k_n)} \}
\]

Because the distance scale associated with the metric is rescaled
along with the transverse coordinates in the rescaling of
\cite{Seiberg-DLCQ}, the rescaling of the partial derivatives in these
expressions cancel the scaling of the moment indices.  Thus,
the higher moment terms which remain in the $R_s
\rightarrow  0$ limit are precisely those corresponding to 0th moments
which remain in the limit.  The only terms in (\ref{eq:IIA-rewritten})
which remain in the limit other than the leading divergent D0-brane
energy term should be finite terms corresponding to the matrix theory
action (\ref{eq:general-background}).  All terms which are linear in
the background and carry positive powers of $R/R_s$ in the limit must
cancel for the IIA action to agree with the matrix theory action.
This gives a number of restrictions on the parts of the IIA currents
with particular scaling dimensions.  The constraints arising in this
fashion for the integrated (monopole) currents are
\bea
({1 \over 2} I_h^{00} + {1 \over 2} I_h^{ii} + {3 \over 2} I_\phi - 
2I_0^0)_0 
&=& 0 \nonumber\\
({1 \over 2} I_h^{00} + {1 \over 2} I_h^{ii} + {3 \over 2} I_\phi - 
2I_0^0)_4
&=& 0 \nonumber\\
({1 \over 2} I_h^{00} + {1 \over 2} I_h^{ii} + {3 \over 2} I_\phi - 
2I_0^0)_8 
&=&T^{--} \nonumber\\
(I_h^{0i}  - I_0^i)_2 &=& 0\nonumber\\
(I_h^{0i}  - I_0^i)_6 &=& T^{-i}\nonumber\\ 
(I_h^{ij})_0 &=& 0\label{eq:constraints}\\
(I_h^{ij})_4 &=&  T^{ij}\nonumber\\
({3 \over 2} I_h^{00} - {1 \over 2} I_h^{ii} - {3 \over 2} I_\phi )_0 
&=& 0\nonumber\\
({3 \over 2} I_h^{00} - {1 \over 2} I_h^{ii} - {3 \over 2} I_\phi )_4 
&=& 
2T^{+-}\nonumber\\
( I_h^{0i} + I_0^i)_2 &=& 2T^{+i}\nonumber\\
({1 \over 8} I_h^{00} + {1 \over 8} I_h^{ii} + {3 \over 8} I_\phi + {1 
\over 
2} I_0^0 )_0 &=& T^{++}  \nonumber
\eea
We will assume that the degrees at which a given current $I_x$ has
nonvanishing contributions are the same as in the $N = 1$ case.  These
are the terms for which we have explicitly written constraints in
(\ref{eq:constraints}).  This assumption agrees with what we know
about the nonabelian Born-Infeld action.  If this assumption is
incorrect, there may be additional terms appearing in the IIA currents
at other orders which do not contribute to the matrix theory action.
Identical relations to (\ref{eq:constraints}) must hold for the
quantities coupling to higher order terms in the Taylor expansion of
the metric. 

Solving the constraints (\ref{eq:constraints}), we find 
\bea
I_h^{00} &=& T^{++} +  T^{+-} + (I_h^{00})_8 + {\cal O} (v^6) 
\nonumber\\
I_h^{0i}
&=&T^{+i} + T^{-i} + {\cal O} (v^5) \nonumber\\
I_h^{ij} &=& T^{ij} +
(I_h^{ij})_8 + {\cal O} (v^6) \label{eq:result-h}\\
I_\phi &=& T^{++} - {1 \over 3} T^{+-} - {1
\over 3} T^{ii} + (I_\phi)_8 + {\cal O} (v^6) \nonumber\\
I_0^0 &=& T^{++} \nonumber\\ I_0^i &=& T^{+i}  \nonumber
\eea
Here, we have  assumed that the one-form field components
$C_0$ and $C_i$ should couple to the D0-brane charge $N/R=T^{++}$ and
spatial current ${\rm Tr}\;(\dot{X}^i)/R=T^{+i}$ so that the last two
expressions are exact.  The fourth order quantities $(I_h^{00})_8$,
$(I_h^{ij})_8$, and $(I_\phi)_8$ are not completely determined by our
analysis, but they must obey the relation
\[
({1 \over 2} I_h^{00} + {1 \over 2} I_h^{ii}  + {3 \over 2} I_\phi)_8 = 
T^{--}
\]
From the $N = 1$
results (\ref{eq:n1}) we expect that
\begin{eqnarray*}
 (I_h^{00})_8 & = &  \frac{3}{2} T^{--}+  (I_h^{00})_{8c} \nonumber\\
 (I_h^{ii})_8 & = &  2  T^{--}+  (I_h^{ii})_{8c} \nonumber\\
 (I_\phi)_8 & = &  -\frac{1}{2} T^{--}+  (I_\phi)_{8c}\nonumber
\end{eqnarray*} 
where $(I)_{8c}$ are quantities of order $v^4$ which contain
commutators or fermions and which vanish in the $N = 1$ case of a
single spinless graviton considered in the previous subsection.

Additional information about the currents should follow from the 
conservation of the D0 brane stress-energy tensor $I_h^{\mu 
\nu}(\vec{x})$ which is defined in terms of the moments $I_h^{\mu \nu 
(k_1 \cdots k_n)}$. As discussed in \cite{mvr}, the relation $D_\mu 
I_h^{\mu \nu} = 0$ implies 
\[
\partial_t I_h^{0 \mu (k_1 \cdots k_n)} = I_h^{k_1 \mu (k_2 \cdots k_n)} 
+ \dots + I_h^{k_n \mu (k_1 \cdots k_{n-1})}
\]
In particular, $(I_h^{ij})_8$ should be precisely determined by 
\[
(I_h^{ij})_8 = (\partial_t T^{+i(j)} + \partial_t T^{-i(j)})_8.
\]

So far, we have only dealt with the case of a background metric $h$,
dilaton field $\phi$ and R-R 1-form field $C$.
The same sort of analysis can be applied for backgrounds
having nonvanishing 2-form $B$ or 3-form $C^{(3)}$ fields and their
duals, and in fact the analysis is simpler in these cases.  
In order to describe nontrivial background antisymmetric tensor
fields, we must generalize the relations (\ref{eq:relations}) which
connect the IIA background fields to the 11D background 3-form field
through the Seiberg limit.  The  $B$ and $C^{(3)}$ fields are related
to components of the 3-form field through
\begin{eqnarray*}
B_{0i} & = &  \hat{A}_{10 \; 0 i} = A_{+ -i}\\
B_{ij} & = &  \hat{A}_{10 \; ij} = \frac{\alpha}{\sqrt{2}}  A_{+ ij} -
\frac{1}{ \alpha \sqrt{2}}  A_{-ij}\\
C^{(3)}_{0ij} & = &  \hat{A}_{0 \; ij} = \frac{\alpha}{\sqrt{2}}  A_{+ 
ij}  + \frac{1}{ \alpha \sqrt{2}}  A_{-ij}\\
C^{(3)}_{ijk} & = &  \hat{A}_{ijk} = A_{ijk}
\end{eqnarray*}
The constraints on the string and D2-brane currents analogous to
(\ref{eq:constraints}) are then
\begin{eqnarray}
(I_2^{ijk})_0 = (I_s^{0i})_0 = (3I_2^{0ij} -I_s^{ij})_2 & = & 0 
\nonumber\\
(I_s^{0i})_4 & = & 3J^{+ -i}\nonumber \\
(I_2^{ijk})_4 & = &  J^{ijk}\label{eq:constraints-a}\\
(3 I_2^{0ij} + I_s^{ij})_2 & = & 6  J^{+ 
ij} \nonumber\\
(3I_2^{0ij} -I_s^{ij})_6 & = & 3  J^{-ij} \nonumber
\end{eqnarray}
from which we can determine
\begin{eqnarray}
I_s^{0i} & = & 3J^{+ -i} +{\cal O} (v^4)\nonumber \\
I_2^{ijk} & = & J^{ijk}+{\cal O} 
(v^4)\label{eq:solution-a}\\
I_2^{0ij} & = &  J^{+ ij} + (I_2^{0ij})_6 +{\cal O} (v^5)
\nonumber\\
I_s^{ij} & = & 3 J^{+ ij} + (I_s^{ij})_6 +{\cal O} 
(v^5)\nonumber
\end{eqnarray}
where the terms $(I)_6$ on the last two lines must satisfy the final
relation in (\ref{eq:constraints-a}). In addition, conservation 
relations for $I_s$ suggest that 
\[
\partial_t I_s^{0i(j)} = I_s^{ji} = - I_s^{ij}
\]
from which it follows that 
\[
(I_s^{ij})_6 = -3\partial_t J^{+-i(j)} = -3 J^{-ij}, \; \; \; \; \; 
(I_2^{0ij})_6 = 0
\]

There are a number of comments worth making about the identifications
(\ref{eq:solution-a}).  First, note that the factor of 3 appearing in
the currents  $I_s$ arises from our somewhat unconventional choice of
normalization for the couplings $A_p J^p$ between a $p$-form field and
its associated current.  Often, a factor of $1/p!$ is included in the
definition of this coupling.  With that redefinition of the currents,
the factors of $3$ in our relations would disappear.  We have chosen
our convention for the coupling to conform with previous literature on
the subject.

Next, we briefly discuss the physical interpretation of the leading
terms in (\ref{eq:solution-a}).
The leading term $J^{+ ij}$ in $I_2^{0ij}$
is the total membrane charge of the D0-brane system.
This result is the T-dual of the statement that $\int F$ on a $p$-brane 
is
the total $(p-2)$-brane charge coupling to the $(p -1)$-form R-R field
\cite{Douglas}.  Although one might think that this should be the only
contribution to the D2-brane charge of the system, 
additional contributions may arise from the geometry of the brane 
embedding
\cite{bvs,ghm,Cheung-Yin,Minasian-Moore}.
Note that while for finite $N$ the integrated membrane charge
$J^{+ ij} = {\rm Tr}\;[X^i, X^j]$ vanishes identically (since a finite
size system can have no net membrane charge) the higher moments of
the D2-brane charge can be nonvanishing and will couple to the
derivatives of the $C^{(3)}$ field.

The leading term in $I_s^{0i}$ is the net string winding charge in
direction $i$; this is simply the T-dual of the Poynting vector giving
momentum on a dual D-brane.  The leading term in the current
$I_s^{ij}$ is perhaps somewhat surprising.  Although this term itself
vanishes for finite size matrices, as mentioned above, the first
moment is nonvanishing.  The existence of this term indicates that
there will be a coupling in the multiple D0-brane action of the form
\begin{equation}
\partial_{[i} B_{jk]} {\rm Tr}\; (X^{[i} X^j X^{k]} + {\rm fermions}).
% \label{eq:}
\end{equation}
We are rather confused as to the physical origin of this term.
Indeed, this term plays a puzzling role in several related situations.
For example, after compactification on $T^3$, the term $J^{+ ij}$
should be related by a duality transformation to the NS5-brane charge
of the IIA theory \cite{grt}, which we discuss below.  It may be
possible to understand the role of this term in the theory by studying
a T-dual system such as a dual multiple D3-brane configuration.  We
discuss the connection with the dual theory briefly in the last
section of this paper.

Now let us consider the currents coupling to the dual fields
$\tilde{B}$ and $\tilde{C}^{(3)}$.  These currents can be derived in a
fashion precisely analogous to the above argument by considering the
fields related to the dual 6-form $\tilde{A}$ of the 11D theory.  We 
find
\begin{eqnarray}
I_4^{0i jkl} & = & 6 M^{+ -i jkl}+{\cal O} (v^4)\nonumber \\
I_5^{ijklmn} & = &  M^{ijklmn}+{\cal O} 
(v^4)\label{eq:solution-ad}\\
I_5^{0ij klm} & = &  M^{+ ij klm} + (I_5^{0ij klm})_6 +{\cal O} (v^5)
\nonumber\\
I_4^{ij klm} & = & 6 M^{+ ij klm} + (I_4^{ij klm})_6
+{\cal O} (v^5)\nonumber
\end{eqnarray}
where
\[
6 (I_5^{0ij klm})_6  -
(I_4^{ij klm})_6= 6M^{-ijklm}.
\]
Just as for $I_s$, conservation relations for $I_4$ suggest that 
\[
\partial_t I_4^{0ijkl(m)} = I_4^{ijklm} 
\]
from which it follows that 
\[
(I_4^{ijklm})_6 = 6\partial_t M^{+-ijkl(m)} = -6 M^{-ijklm}, \; \; \; \; 
\; (I_5^{0ijklm})_6 = 0.
\]

The leading term in $I_4^{0ijkl}$ is the net D4-brane charge
\cite{bss,grt}.  This is the dual of the instanton number on a
D$p$-brane.  Unfortunately, we only know from matrix theory the
components of the 5-brane current $M^{-IJKLM}$ with one $-$ index.
Thus, we can only determine the leading term in the components
$I_4^{0ijkl}$ of the D4-brane current, and we have no information
about the leading terms in the remaining components of $I_4$
or any components
of the NS5-brane current $I_5$.  The absence of any known operator for
the transverse 5-brane charge $M^{+ ijklm}$ in matrix theory is an
infamous problem.  No operator of this form appears in the
supersymmetry algebra \cite{bss} or in the one-loop effective
potential \cite{Dan-Wati-2}.  Nonetheless, we should expect higher
moments of this operator to appear, corresponding to local transverse
5-brane charge.  It has been argued that in a T-dual 3-brane picture
the desired operator is S-dual to the charge of a D5-brane
perpendicular to the 3-brane \cite{grt} (described by the operator
$J^{+ ij}$ mentioned above), although no explicit description of this
dual operator has been given.  It would be nice to have a better
understanding of these terms in the multiple D0-brane action.

Finally, we consider the currents coupling to the dual field
$\tilde{C}$.   We expect the IIA 6-brane current to couple to this
field.  Unlike the other branes whose currents we have considered, the
IIA 6-brane does not arise in a simple fashion from the dimensional
reduction of the membrane or the 5-brane of 11D M-theory.  Rather, the
IIA 6-brane represents a nontrivial metric of Kaluza-Klein monopole
form in the 11D theory.  Nonetheless, in \cite{Mark-Wati-3} we found a
matrix theory description of interactions between such metrics and
0-branes.  This appeared in the form of a term in the 2-body matrix
theory potential which coupled the 0-brane stress tensor to a 10D
6-brane current $S^{\mu \nu \rho \lambda \sigma \tau \upsilon}$.
Since this current already has an essentially 10D form, it is natural
to map it directly to the 6-brane current we expect in the IIA
theory.  Thus, we believe that the 6-brane current of a system of many
0-branes which couples to the background $\tilde{C}$ field will be
given by
\begin{eqnarray}
I_6^{0ijklmn} & = &  S^{+ijklmn} +{\cal O} (v^5) \label{eq:solution-6}\\
I_6^{ijklmn p} & = &  S^{ijklmn p}+{\cal O} (v^6)\nonumber
\end{eqnarray}
where the matrix theory form of the 6-brane current is given in the 
Appendix.

\subsection{Summary of results for multiple D0-brane action in IIA}

We summarize here our results for the terms in the multiple D0-brane
action which couple linearly to the background fields of type IIA
supergravity and their derivatives.  The full action including all
linear terms
is given by (\ref{eq:IIA-general})
 \bea
S  &= &  S_{{\rm flat}}  +
 \int dt  \sum_{n=0}^{\infty} {1 
\over n!} \left[
\frac{1}{2}
(\partial_{k_1}\cdots 
\partial_{k_n} h^{IIA}_{\mu \nu}) \; I_h^{\mu \nu (k_1 \cdots k_n)}
+ (\partial_{k_1}\cdots \partial_{k_n} \phi) 
\; I_{\phi}^{(k_1 \cdots k_n)}\right.\nonumber\\ 
& & \hspace{1in}+ (\partial_{k_1}\cdots 
\partial_{k_n} C_{\mu }) \; I_0^{\mu (k_1 \cdots k_n)}
+
%\frac{1}{7!}
 (\partial_{k_1}\cdots 
\partial_{k_n} \tilde{C}_{\mu \nu \lambda \rho \sigma \tau \zeta }) 
\; I_6^{\mu \nu \lambda
\rho \sigma \tau \zeta (k_1 \cdots k_n)}
\nonumber\\
& & \hspace{1in}+
%\frac{1}{2}
 (\partial_{k_1}\cdots \partial_{k_n} 
B_{\mu \nu}) \; I_s^{\mu \nu (k_1 \cdots k_n)} 
+ 
%\frac{1}{6!}
(\partial_{k_1}\cdots \partial_{k_n} 
\tilde{B}_{\mu \nu \lambda \rho \sigma \tau}) \; I_5^{\mu \nu \lambda
\rho \sigma \tau
 (k_1 \cdots k_n)} \nonumber\\
& &\hspace{1in} \left.+ 
%\frac{1}{6}
(\partial_{k_1}\cdots 
\partial_{k_n} C^{(3)}_{\mu \nu \lambda }) \; I_2^{\mu \nu \lambda (k_1 
\cdots 
k_n)}
+
%\frac{1}{5!}
 (\partial_{k_1}\cdots \partial_{k_n} \tilde{C}^{(3)}_{\mu \nu \lambda
\rho \sigma }) \; I_4^{\mu \nu \lambda \rho \sigma (k_1 \cdots
k_n)}\right] \nonumber \eea The multipole moments of the stress tensor
$I_h$ and currents coupling to the background dilaton and R-R 1-form
field have leading terms given by (\ref{eq:result-h}).  The currents
coupling to the NS-NS antisymmetric 2-form field and the R-R 3-form
field have leading terms given by (\ref{eq:solution-a}).  Of the
currents coupling to the duals of these two fields, we have only been
able to identify leading term in the the moments of the 4-brane
current component $I_4^{0ijkl}$ as described in
(\ref{eq:solution-ad}).  We believe that the leading terms in the
components of the 6-brane current coupling to the dual of the 1-form
field are as given in (\ref{eq:solution-6}).

We have derived these results based on our proposal in
\cite{Mark-Wati-3} for the form of the matrix theory action in weak
background fields and Seiberg's scaling argument.  Our results agree
with the known terms in the $N = 1$ Born-Infeld action, and with the
known BPS charges of the multiple D0-brane system.  In the following
section we give a simple test of the results for the terms coupling to
the background metric.  Further possible tests, applications, and
extensions of these results are discussed in the concluding section.

\section{Tests of the action}

In this section, we test our proposals for the general background 
actions through two related calculations. First, we consider the 
D0-brane action in a background describing two seperated branes in a 
curved space ($h_{ij} \ne 0$). We determine the masses of off diagonal 
bosonic and fermionic fields and show that these exactly match the 
geodesic distance to leading order in $h$ in agreement with the 
constraint suggested by Douglas. Next, we consider the analogous 
background in matrix theory and compute the leading order one-loop 
potential between two gravitons in a curved transverse space, showing 
that curved-space supergravity predictions are reproduced.

\subsection{The geodesic length criterion}
\label{sec:masses}

One of the earliest discussions of the problem of formulating a
low-energy theory for many D0-branes moving in a curved space was given 
 in \cite{Douglas-curved}.  In that paper, Douglas argued that one of 
the most basic conditions which such an action must satisfy is that in a
background corresponding to a pair of D0-branes living at points $x$
and $y$ there should be light fields with masses equal to the geodesic
length between these points.  This condition, together with additional
axiomatic assumptions, was used by Douglas, Kato and Ooguri in
\cite{dko} to give the first few terms in the D0-brane action on a
general Calabi-Yau 3-fold which preserves some supersymmetry. In this 
section we show that our formulation of the linearized coupling
to a weak background in the multi-D0-brane action satisfies Douglas's 
geodesic length criterion.

\subsubsection{Setup}

We wish to consider a pair of D0-branes at separated points in a weakly 
curved space. We assume that the transverse metric is described by a 
small perturbation about a flat background, $g_{ij} = \delta_{ij} + 
h_{ij}$, while the other components of the metric and the other 
background fields are trivial. Without loss of generality, we choose 
coordinates so that one brane is at the origin, while the other has 
transverse coordinates $r^i$. The situation is described by the 
multi-D0-brane action (\ref{eq:IIA-general}) with non-zero $h_{ij}$ and 
fields expanded about background matrices as
\begin{eqnarray}
X^i &=  &\left(\begin{array}{cc}
 r^i & 0\\
0 & 0
\end{array} \right)+
\left(\begin{array}{cc}
 \zeta^i & z^i\\
\bar{z}^i & \tilde{\zeta}^i
\end{array} \right)
\label{eq:background}
\\
\Theta & = &\left(\begin{array}{cc}
 0 & 0\\
0 &   0
\end{array} \right)+
\left(\begin{array}{cc}
 \eta^i & \chi^i\\
\bar{\chi}^i &  \tilde{\eta}^i
\end{array} \right), \nonumber
\end{eqnarray}
Here, the fields $\zeta$, $z$, $\eta$, and $\chi$ represent fluctuations 
about the background. 
   The geodesic length condition formulated by Douglas states that the 
masses of the off-diagonal fields $z$ and $\chi$, which arise from 
strings stretched between the separated branes, should precisely match 
the geodesic length measured by the metric $h_{ij}$ between the points 0 
and $r^i$. We will now compute both the geodesic length and the 
oscillator masses and show that they agree.
   
\subsubsection{Geodesic distance}

We begin with the geodesic length between points $0$ and $r^i$.  This
geodesic length is the minimum value of the length
\begin{equation}
\int_\gamma ds = \int_0^{r^i} \sqrt{g_{ij}dx^i dx^j}
\label{eq:geodesic-length}
\end{equation}
taken over all paths $\gamma$ between the two points.  Because the 
geodesic path is an extremum of this functional, the variation of the 
length under a small variation $\delta \gamma$ of the path is of order
$(\delta \gamma)^2$.  Since we are interested in changes in the length
which are linear in the background metric, we can therefore neglect
effects from the change of the geodesic path and simply evaluate the
change in the geodesic length by integrating (\ref{eq:geodesic-length}) 
along the straight line which is the geodesic in the flat metric.
Thus, we take
\[
x^i(\lambda) = r^i \lambda 
\]
and find
\beas
d(0,r^i) &=& \int_0^1 d\lambda \sqrt{g_{ij}(\vec{x}(\lambda))\dot{x}^i 
\dot{x}^j}\\
&=& \int_0^1 d\lambda \sqrt{r^2+h_{ij}(\vec{r}\lambda)r^i r^j}\\
&=& \int_0^1 d\lambda \{r + {1 \over 2r} h_{ij}(\vec{r}\lambda)r^i r^j  
+{\cal O} (h^2)\}\\
&=& r + {1 \over 2r} r^i r^j H_{ij}
\eeas
where 
\beas
H_{ij} &=& \int_0^1 d  \lambda \, h_{ij}(\lambda \vec{r})\\
&=& \sum_{n=0}^\infty{1 \over (n+1)!} (r \cdot \partial)^n h_{ij}(0)
\eeas

This gives us the geodesic length between the two points to
linear order in the background metric. In the following section, it will 
be most convenient to compare squared oscillator masses with the squared 
geodesic length, given by
\be
\label{eq:length}
d^2 = r^2 + r^i r^j H_{ij} +{\cal O} (h^2)
\ee

\subsubsection{Oscillator masses}
\label{sec:oscillator-masses}

We now calculate the masses of the off-diagonal fields. From
(\ref{eq:IIA-general}), we find that the $N=2$ D0-brane action in
the case of a transverse background metric has leading terms \bea S
&=& {1 \over 2R} \int dt \; {\rm Tr}( D_t X^i D_t X^i + {1 \over 2}
[X^i, X^j] [X^i, X^j] + i \Theta D_t \Theta - \Theta[\slash{X},
\Theta]) \nonumber\\ & & + {1 \over 2} \int dt \; \sum_{n=0}^\infty {1
\over n!} \partial_{k_1} \cdots \partial_{k_n} h_{ij} T^{ij(k_1 \cdots
k_n)}
\label{eq:Trans-action}
\eea
Here, 
\bea
T^{ij(k_1 \cdots k_n)} &=& {1 \over R} {\rm STr}\;
\left( \left\{  D_t X^i D_t X^j - 
[X^i, 
X^k] [X^k, X^j] \right.\right. \label{eq:tij}\\
& &\left.\left.\hspace{1in} 
- {1 \over 4} \Theta \gamma^i [X^j, \Theta] - {1 \over 
4} \Theta \gamma^j [X^i, \Theta] \right\} X^{k_1} \cdots X^{k_n}\right)
\nonumber\\ & 
& 
\hspace{0.3in} + \tilde{T}^{ij(k_1 \cdots k_n)}\nonumber
\eea
We note here that this is precisely the action for Matrix theory in a 
background $h_{ij}$ to leading order in the metric. In the next section 
we will use exactly this action to calculate the one-loop potential 
between two gravitons, taking the same background (\ref{eq:background}), 
though allowing 
$\vec{r}$ to be a function of time. Such a calculation is simplest using 
a gauge fixed version of the action in which we choose the background 
field gauge, adding a term
\[
S_{fix} = {1 \over R} \int (-D_t X^0 + i[B^i, X^i])^2
\]
plus the appropriate ghost terms. For later convenience, we will analyze 
this gauge fixed version, keeping in mind both the Matrix theory 
interpretation and D0-brane interpretations. 

Unlike the Matrix theory action, the complete D0-brane action contains 
additional terms both in the background intependent part and coupled to 
$h_{ij}$. However these cannot contribute to the quadratic action since 
they contain more than two matrices (eg $\dot{X}^i$, $F_{ij}$, $\Theta$) 
in which there are no entries depending only on background fields. 
Similarly, the terms  $\tilde{T}^{ij(k_1 \cdots k_n)}$, whose form has 
not been worked out for $n>1$ involve at least two fermions $\Theta$ and 
one power of $\dot{X}$ or $[X^i, X^j]$ and so contribute only cubic and 
higher order terms to the action, irrelevant for determining the 
oscillator masses or computing the one loop potential.

   We now replace $X$ and $\Theta$ in the action with the matrices given
in (\ref{eq:background}), and write down the terms in the action 
quadratic in the off
diagonal fields $z$ and $\chi$.  It turns out that the symmetrization
prescription for ordering the matrices in $T^{ij(k_1 \cdots k_n)}$ is
very important here, since most of the orderings give no contribution to
the quadratic terms we are interested in.  For example, the first term
in (\ref{eq:tij}) contains a term
\beas
\lefteqn{{1 \over 2 n!} (\partial_{k_1} \cdots \partial_{k_n} h_{ij}) \; 
\; 
{\rm STr}\;(\dot{X^i} \dot{X^j} \; X^{k_1} \cdots X^{k_n})} \\
&=& {1 \over 2 n!} (\partial_{k_1} \cdots \partial_{k_n} h_{ij})\; \; 
{1 \over 
n+1} 
\sum_{m=0}^n {\rm Tr}\;(\dot{X}^i X^{k_1} \cdots X^{k_m} \dot{X}^j 
X^{k_{m+1}} 
\cdots X^{k_n})
\eeas
for which only the $m=0$ and $m=n$ terms contribute to the quadratic 
action in the off diagonal field $z$. Summing over $n$, this 
contribution gives
\beas
& &\sum_{n=0}^\infty {1 \over (n+1)!} (\partial_{k_1} \cdots 
\partial_{k_n} 
h_{ij}) (\dot{\bar{z}}^i \dot{z}^j ) r^{k_1} \cdots r^{k_n}\\
&=&\dot{\bar{z}}^i \dot{z}^j \sum_{n=0}^\infty {1 \over (n+1)!} (r 
\cdot 
\partial)^n h_{ij} 
\\
&\equiv&H_{ij} \dot{\bar{z}}^i \dot{z}^j
\eeas
Note that the quantity $H_{ij}$ is simply the function $h_{ij}$ 
integrated over the straight line trajectory between $0$ and $\vec{r}$,
\[
H_{ij} = \int_0^1 h_{ij}(\lambda \vec{r}) d \lambda
\]
which also appeared at first order in the geodesic distance formula 
(\ref{eq:length}). Using this definition, it is straightforward to write 
down the remaining terms in the quadratic actions for each of the off 
diagonal fields.

\vspace{0.15in}

\noindent {\bf Bosonic Terms}

\vspace{0.15in}

In exactly the same way as for the terms just derived, we find that the 
complete set of terms to leading order in $h$ for the nine transverse 
bosonic 
fields is
\[
S_B = -\bar{z}^i\{(\partial_t^2 + r^2) (\delta_{ij} + H_{ij}) + r^k r^l 
H_{kl} \delta_{ij} - r^i r^k H_{kj} - H_{ik} r^k r^j \}z^j     
\]
Note that the matrix $(\delta_{ij} + H_{ij})$ multiplies all terms not 
containing $h$. Since the remaining terms are already of order $h$, if 
we 
redefine $z^i$ to eliminate this factor in the first terms, the 
remaining 
terms will only be changed at second order in $h$. After such a field 
redefinition, the kinetic term is proportional to the identity, and the 
squared oscillator masses are therefore given by the eigenvalues of the 
constant matrix
\beas
{\bf M} &=& r^2 \left\{ (1+H_{rr})\identity_{9 \times 9}
+ \left( \begin{array}{cccc} 0 & \cdots & 0 & H_{1r}\\ \vdots & & & 
\vdots \\
0 & & 0 & H_{8r} \\ H_{1r} & \cdots & H_{8r} & -2H_{rr} \end{array} 
\right) + 
{\cal O} (h^2) \right\}
\eeas
Here, to simplify the formulae, we have made a rotation so that 
$\vec{r}$ lies in the $9$ direction, which we refer to using the index 
$r$. It is straightforward to solve directly for the eigenvalues and 
eigenvectors of this mass matrix, and one finds to this order that the 
masses are
\[
m_1^2 = \cdots = m_7^2 = r^2(1 + H_{rr}), \; \; m_8^2 = r^2(1 + 
\sqrt{H_{rr}^2 
+ 
H_{ri} H_{ir}}), \; \; m_9^2 = r^2(1 - \sqrt{H_{rr}^2 + H_{ri} H_{ir}})
\]
where the index $i$ is summed from 1 to 8. The oscillators with masses 
$m_9$ and $m_8$ correspond to directions lying in the plane defined by 
the $r$ direction and the perpendicular vector $H_{ir}$. The remaining 
oscillators correspond to the directions perpendicular to these and have 
masses which precisely match the geodesic distance (\ref{eq:length}) to 
leading order 
in $h$. 
The agreement between the masses of these perpendicular oscillators
and the geodesic distance is precisely the criterion used by Douglas
et al.
in \cite{Douglas-curved,dko}  to constrain the leading terms in
multiple D0-brane action on certain classes of manifolds.  The fact
that the oscillators corresponding to fields not perpendicular to the
separation have different masses is also expected.  In the non-gauge
fixed theory, the off-diagonal fields in the direction of the
separation between the branes simply give a combination of a gauge
rotation of the system and a relative motion of the D0-branes along a
flat direction.  This effect explains the failure of the masses $m_8,
m_9$ to satisfy Douglas's criterion.

\vspace{0.15in}

\noindent {\bf Gauge Field}

\vspace{0.15in}

  For a time independent $\vec{r}$, there is no mixing between gauge 
field and the other bosonic oscillators in the quadratic action, and we 
find that the quadratic terms involving the off diagonal field $z_0$ are 
simply
\[
S_A =-\bar{z}_0 \{ \partial_t^2 + r^2 + r^2H_{rr} \} z_0
\]
Hence, the off diagonal gauge field also has mass equal to the geodesic 
distance to leading order in $h$,
\[
m_0^2 = r^2 + H_{rr} \,.
\]

\vspace{0.15in}

\noindent {\bf Ghost Fields}

\vspace{0.15in}

Since our gauge fixing term does not depend on the background 
metric, the off diagonal ghost fields will have a mass given by $m_g^2 = 
r^2$, as in the flat space case, with action
\[
S_G = -\bar{c} \{ \partial_t^2 + r^2 \} c
\]

\vspace{0.15in}

\noindent {\bf Fermionic Fields}

\vspace{0.15in}

Proceeding in the same way for the quadratic fermion action, we find 
that the action quadratic in the off diagonal field $\chi$ is
\[
S_F =-\bar{\chi}_\alpha \{i\partial_t + \gamma^i_{\alpha \beta} (r^i + 
{1 \over 
2} H_{ij} r^j) \} \chi_\beta
\]
Thus, in the presence of a background metric, the quadratic fermion 
action is only changed by a shift
\[
r^i \rightarrow r^i + {1 \over 2} H_{ij} r^j
\]
so the sixteen fermion fields $\chi$ have a mass squared matrix given 
by 
\[
M_f^2 = \identity (r^2 + H_{ij} r^ir^j + {\cal O} (h^2)) = \identity 
r^2(1 + 
H_{rr} + 
{\cal O} (h^2))
\]
We see that all the fermionic oscillators have a mass which reproduces 
the geodesic distance to leading order in $h$.

\vspace{0.15in}

\noindent {\bf Summary}

\vspace{0.15in}

To summarize, we have found eight complex bosons (including the gauge 
field) and sixteen real fermions with masses equal to the geodesic 
distance between $0$ and $\vec{r}$. Additionally, there are complex 
bosons with $m^2 = r^2(1 + \sqrt{H_{rr}^2 + H_{ri} H_{ir}})$ and $m^2 = 
r^2(1 - \sqrt{H_{rr}^2 + H_{ri} H_{ir}})$ and two complex ghosts with 
$m^2 = r^2$.  
  
   Thus, the sum of the squared masses weighted by the number of 
degrees of freedom is identical for the fermions and the bosons 
(including the ghosts with negative weight), and for both sets of 
fields, the average mass squared per degree of freedom is exactly the 
geodesic distance (\ref{eq:length}). We have now seen that the geodesic 
distance 
criterion is precisely realized in the proposed multi-D0-brane action.

\subsection{Graviton interactions}

In this section, we use the general background Matrix theory action to 
study the interactions between two gravitons with unit momentum around a 
lightlike circle in a weakly curved space. In various cases, we will 
compute the one-loop effective action to first order in the metric and 
compare with the interactions expected from DLCQ supergravity with a 
curved background.

The relevant matrix theory action and background are exactly the same as 
those considered for D0-branes in the previous section. In this case, 
however, we do not wish to restrict to gravitons which are fixed in the 
transverse space, so we allow $\vec{r}$ to be a function of time.

As for the case of flat space Matrix theory, we should require that
our background matrices satisfy the equations of motion.  For block
diagonal backgrounds, the equations of motion decouple for each block,
so for our case, we require that $\vec{r}(t)$ satisfy the equations of
motion derived from the $U(1)$ action (\ref{graviton}).  For a metric 
which is non-trivial only in the transverse directions, the equations of 
motion are
\be
\label{eq:eom}
\ddot{r}^i = \dot{r}^k \dot{r}^l g^{im} (\vec{r}) \{ {1 \over 2} 
\partial_m 
g_{kl} (\vec{r}) - \partial_k g_{ml} (\vec{r})  \}
\ee  
  
This is just the equation for a free non-relativistic particle moving in
a curved space.  We will consider two simple cases of trajectories which
trivially satisfy these equations of motion.  First, we may consider the
static case $\vec{r}(t) = \vec{r}(0)$.  The second case is one for which
the metric has a flat direction $i$ (so that $h_{ij} =
\partial_ig_{jk} = 0$)
and we take the particle to have some velocity in this direction.  In 
this case, the right hand side of (\ref{eq:eom}) vanishes, so that 
$r^i(t) = r^i + v^it$.

\subsubsection{Supergravity predictions}

Before proceeding with the matrix theory calculation, we would like to 
see what supergravity predicts for the interaction potential between two 
gravitons in a weakly curved space. 

As above, we start with a static metric $g_{ij}(\vec{x}) = \delta_{ij} + 
h_{ij}(\vec{x})$ which is assumed to satisfy the source-free Einstein 
equations   
\be
\label{eq:Ricci}
0 = R_{ij} ={1 \over 2}(\Delta h_{ij} - \partial_i \partial_k h_{kj} - 
\partial_j \partial_k h_{ki} + \partial_i \partial_j h_{kk}) + {\cal 
O}(h^2)
\ee
We may find the potential between a pair of gravitons in this space by 
treating one as a source for a perturbation about the metric $g$ and 
reading off the potential from the probe action (\ref{graviton}), 
keeping only terms arising from the perturbation in the original metric 
due to the presence of the source graviton. 

For our source, we choose the graviton which sits at the origin of the 
transverse space with unit momentum in the compact direction. The 
stress-energy tensor for this particle still has only a single 
non-vanishing component, 
\[
T^{++} = T_{--} = {1 \over 2 \pi R^2} \delta(\vec{x}), 
\]
The presence of this graviton will result in a perturbation of the 
metric $g_{ij}$ which we denote by $\gamma_{ij}$. The fact that $T_{--}$ 
is the only non-vanishing component of the stress-energy tensor 
simplifies things considerably, and as with the flat space case, we may 
solve the Einstein equation taking only the component $\gamma_{--}$ to 
be non-zero. In this case, the condition that the perturbed metric $g + 
\gamma$ should continue to satisfy the Einstein equation with source $T$ 
reduces at leading order in $\gamma$ to the covariant Laplace equation,
\be
g^{ij}\nabla_i \nabla_j \gamma_{--}  =  g^{ij} (\partial_i \partial_j - 
\Gamma^k_{ij} \partial_k ) \gamma_{--} = 2\kappa_{11}^2 T_{--}
\label{eq:laplace}
\ee
(see, for example \cite{Weinberg}). We are only interested in the 
solution at leading order in $h$ (the original background metric), so we 
expand
\[
\gamma_{--} \equiv \gamma_0 + \gamma_1 + {\cal O}(h^2)
\]
Here $\gamma_0$ is the part independent of $h$, equal to the flat space 
solution (ignoring non-numerical constants)
\be
\label{eq:flat}
\gamma_0 = {15 \over 2 r^7}
\ee
while $\gamma_1$ is the part linear in $h$. 

In the case where $g_{ij}$ is the metric corresponding to some choice of 
coordinates on flat space, the exact solution to (\ref{eq:laplace}) must 
be given by a covariant version of (\ref{eq:flat}), replacing $r$ with 
the geodesic distance $d$ between $0$ and $r$. In this case, using 
(\ref{eq:length}), we have 
\be
\label{eq:hlinear}
\gamma = {15 \over 2} \{ {1 \over r^7} - {7 r^i r^j \over 2r^9} \int_0^1 
d \lambda h_{ij}(\lambda \vec{r})\} + {\cal O}(h^2) 
\ee
In fact, this solves (\ref{eq:laplace}) to leading order in $h$ in any 
case where the metric $g$ is Ricci-flat as may be verified explicitly by 
substitution. The $h$ independent part of (\ref{eq:laplace}) reads  
\[
\partial^2 \gamma_0 = 2\kappa_{11}^2 T_{--}, 
\]
and is just the statement that $\gamma_0$ is the solution for flat 
space. The linear terms in $h$ in equation (\ref{eq:laplace}) read
\[
\partial^2\gamma_1 = h_{ij}\partial_i \partial_j \gamma_0 + ( \partial_i 
h_{ik} - {1 \over 2} \partial_k h_{ii}) \partial_k \gamma_0
\]
and substituting for $\gamma_0$ and $\gamma_1$ from (\ref{eq:hlinear}) 
it is not hard to check that this holds, making use of the 
Ricci-flatness condition (\ref{eq:Ricci}), the identity
\[
\partial_\lambda h(\lambda \vec{r}) = (r \cdot \partial) h(\lambda 
\vec{r}),
\]
and various integrations by parts. This is done most easily by choosing 
coordinates so that $h$ satisfies the harmonic gauge condition in which
\[
\partial_i h_{ij} = { 1 \over 2} \partial_j h_{ii}, \;\;\;\;R_{ij} = 
\partial^2 h_{ij} + {\cal O}(h^2)
\].

Thus, to leading order in the backgrounds we are considering, the metric 
perturbation due to the presence of a graviton at the origin is simply
\[
\gamma_{--}(\vec{x}) = {15 \over 2 d^7(\vec{x})}
\]
where $d$ is the geodesic distance between 0 and $\vec{x}$. Recalling 
the action (\ref{graviton}) for the probe graviton moving in the metric 
produced by this source, we see that to leading order in the background, 
the curved space graviton-graviton potential is simply
\[
V = -{15 \over 16} {v^4 \over d^7}
\] 
In the next sections, we will carry out the graviton potential 
calculation in Matrix theory to compare with this supergravity 
prediction.

\subsubsection{Static case}
  
    First, we consider the static case in which both gravitons have zero
transverse velocity. In this case, as in flat space, we expect the 
potential to vanish at leading order in the transverse metric, as shown 
above. 

  For this case, we may directly apply the results of section 
\ref{sec:masses}. The complete action quadratic in the off diagonal 
fields is 
\bea
S&=&S_B+S_A+S_F+S_G \nonumber \\   
S_B &=& -\bar{z}^i\{(\partial_t^2 + r^2) (\delta_{ij} + H_{ij})  
\nonumber \\
& &\hspace{0.5in} + (r^k r^l 
H_{kl})\delta_{ij} - r^i r^k H_{kj} - H_{ik} r^k r^j \}z^j  \nonumber \\
S_A &=& -\bar{z}_0 \{ \partial_t^2 + r^2 + r^i r^j H_{ij} \} z_0 
\label{eq:stat-action}\\
S_F &=& -\bar{\chi}_\alpha \{i\partial_t + \gamma^i_{\alpha \beta} (r^i 
+ {1 \over 2} H_{ij} r^j  \} \chi_\beta  \nonumber \\
S_G &=& -\bar{c} \{ \partial_t^2 + r^2 \} c \nonumber 
\eea

As argued above, this leads to eight complex bosons (including the gauge 
field) and sixteen real fermions with masses equal to the geodesic 
distance between $0$ and $\vec{r}$ as well as complex bosons with $m^2 = 
r^2(1 + \sqrt{H_{rr}^2 + H_{ri} H_{ir}})$ and $m^2 = r^2(1 - 
\sqrt{H_{rr}^2 + H_{ri} H_{ir}})$ and two complex ghosts with $m^2 = 
r^2$. 

The vanishing of the one loop potential to leading order in $h$ is 
ensured by the fact that the sum of the squared masses weighted by 
number of degrees of freedom is identical for the fermions and bosons 
(including ghosts weighted by -1). This follows since the one loop 
effective action depends only on the oscillator masses and is given by:
\beas
e^{i\Gamma_{1 loop}}&=&\det{}^{-8}(\partial_t^2 + r^2(1 + H_{rr})) 
\det{}^{8}(\partial_t^2 + r^2(1 + 
H_{rr})) \det{}^{2}(\partial_t^2 + r^2)\\
& &\det{}^{-1}(\partial_t^2 + r^2(1 + \sqrt{H_{rr}^2 + H_{ri} 
H_{ir}}))\det{}^{-1}(\partial_t^2 + r^2(1-\sqrt{H_{rr}^2 + H_{ri} 
H_{ir}}))\\
&=&1 + {\cal O} (h^2)
\eeas

Hence, in the static case, we find agreement with our expectations 
from supergravity.

\subsubsection{Velocity dependent potential}

We now consider the case of two gravitons with relative velocity.  
Here, we assume that the particle with initial position $\vec{r}$ moves 
in a direction $\vec{v}$ which is perpendicular to $\vec{r}$ and in 
which the metric is flat. In this case, we have shown above that 
supergravity predicts a potential 
\[
-{15v^4 \over 16 d^7}
\]
where $d$ is the geodesic separation distance.

  To simplify our calculations, we rotate coordinates so that $\vec{r}$ 
and $\vec{v}$ lie in coordinate directions which we denote by indices 
$r$ and $v$. Thus
\[
g_{vv}(\vec{x}) = 1, \;  g_{vr}(\vec{x})= g_{vi}(\vec{x}) = 
\partial_vg_{ij}(\vec{x}) = 0 
\]
These equations ensure that the matrix theory equations of motion 
(\ref{eq:eom}) 
are satisfied for the trajectory $\vec{r}(t) = \vec{r} + \vec{v}t$ that 
we are considering. Expanding the action (\ref{eq:stat-action}) about 
this background, we 
find that the quadratic action for the off diagonal fields is equal to 
the action for the static case (where $r^i$ is interpreted as 
$r^i(t)=r^i + v^it$) plus extra terms:
\[
S_v = 2i\bar{z}^iv^iz^0 - 2i\bar{z}^0v^iz^i
\]
Note that these terms (in which the background appears as 
$\dot{\vec{r}}$ rather than just $\vec{r}$) come only from the flat 
space action, since in the metric dependent terms, $\dot{X}^i$ only 
appears coupled to $h_{ij}$ while $\dot{r}^i h_{ij} = 0$. As a result, 
the remaining calculation is almost identical to the flat space 
calculation, performed  in \cite{DKPS}.
%for example in \cite{Becker-Becker}. 
To see this, 
we note that the 
complete action in this case may be written (eliminating a factor 
$(\delta_{ij} + H_{ij})$ as above to diagonalize the boson kinetic term) 
\beas
S&=&S_B+S_A+S_v+S_F+S_G\\   
S_B &=& -\bar{z}^i\{(\partial_t^2 + d^2 + v^2t^2) \} z^i\\
& &\hspace{0.5in}  + \bar{z}^r\{r^2H_{ri}\}z^i + 
\bar{z}^i\{r^2H_{ir}\}z^r + \bar{z}^v\{ H_{ri}rvt\}z^i + 
\bar{z}^i\{H_{ir}rvt\}z^v \\
S_A &=& -\bar{z}_0 \{ \partial_t^2 + d^2 + v^2t^2 \} z_0\\
S_v &=& 2i\bar{z}^iv^iz^0 - 2i\bar{z}^0v^iz^i\\
S_F &=& -\bar{\chi}_\alpha \{i\partial_t + \gamma^i_{\alpha \beta} (d^i 
+ v^it )  \} \chi_\beta \\
S_G &=& -\bar{c} \{ \partial_t^2 + d^2 \} c\\
& &\hspace{0.5in} + \bar{c} \{H_{rr}\} c 
\eeas
where we have defined 
\[
d^i = r^i + {1 \over 2} H_{ij} r^j
\]
so that $d^2$ is the squared geodesic distance. Apart from the terms 
with explicit factors of $H$ in the boson and ghost actions, this is 
exactly the flat space action with $r^i$ replaced by $d^i$. Recalling 
the flat space calculation, we see that at zeroth order in $H$, the 
oscillators $z^i$, $i=1,\cdots, 7$, and $z^r$ are degenerate with mass 
$d^2+v^2t^2$ while $z^v$ and $z^0$ combine into oscillators with 
non-degenerate masses $(d^2 + v^2t^2 \pm 2v)$. Adding the perturbation
\be
\label{pert}
\bar{z}^r\{r^2H_{ri}\}z^i + \bar{z}^i\{r^2H_{ir}\}z^r + \bar{z}^v\{ 
H_{ri}rvt\}z^i + \bar{z}^i\{H_{ir}rvt\}z^v
\ee
we note that the last two terms do not affect the masses at leading
order in $h$, since if we change coordinates to diagonalize the
leading order mass matrix, these contribute only to non-diagonal
matrix elements connecting eigenvectors of different mass. (Recall
from basic perturbation theory that for eigenvectors $|A_1\rangle,
\cdots |A_n\rangle$ with degenerate zeroth order eigenvalues and
eigenvectors $|B_1\rangle, \cdots |B_m\rangle$ with non-degenerate
zeroth order eigenvalues that the first order shift in the eigenvalues
for $|B_i\rangle$ come only from the matrix element $\langle
B_i|M|B_i\rangle$, while the first order eigenvalues for the space 
spanned
by $|A_i\rangle$ are determined only by the submatrix $\langle
A_i|M|A_j\rangle$).

   The first two terms in (\ref{pert}) have the same effect as they did 
in the static case, to shift two of the degenerate boson masses by
\beas
\Delta m^2 &=& - H_{rr} \pm \sqrt{H_{rr}^2 + H_{ri}H_{ir}}\\
&\equiv&-H_{rr} \pm G
\eeas
Meanwhile, the perturbation in the ghost action shifts the two ghost 
masses by
\[
\Delta m_g^2 = - H_{rr}
\]
Defining
\[
F(x) = det(\partial_t^2 + (d^2 - x) +v^2t^2)
\]
we find that since all of the mass shifts are time independent, the one 
loop effective action is simply
\beas
e^{i\Gamma} &=& e^{i \Gamma_d} {F(H_{rr})F(H_{rr}) \over F(H_{rr} + 
G)F(H_{rr} - G)}\\
&=& e^{i \Gamma_d}(1 + {\cal O} (h^2))
\eeas
where $\Gamma_d$ is the flat-space potential with $r$ replaced by the 
geodesic length $d$. We conclude that the leading order one loop 
potential is simply given by
\[
V=-{15 \over 16} {v^4 \over d^7}
\]
as predicted by supergravity.

\section{Discussion}
  
In this paper we have derived the leading terms in the multiple
D0-brane action in a weakly curved background metric, dilaton, and
antisymmetric tensor background fields.  This action was derived by
using Seiberg's scaling arguments on an action we recently proposed
for the M(atrix) model of M-theory in weak background fields.  We
found explicit forms for the IIA stress-energy tensor of a multiple
D0-brane system, as well as the components of D2-brane, D4-brane,
D6-brane and fundamental string currents which couple to the
background R-R and B fields of the IIA theory.  We tested our action
by verifying that it satisfies Douglas's geodesic length criterion.
We also showed that the corrections to the one-loop effective
potential between a pair of individual 0-branes correctly reproduce
curved space supergravity results.

The results presented in this paper give for the first time a
systematic description of the linear coupling between a system of
multiple D-branes and background supergravity fields.  There are a
number of previous discussions of this sort of action in the
literature to which our results can be related.  
In \cite{dko}, Douglas, Kato and Ooguri used
Douglas's geodesic length criterion
and other axioms including an
assumption of supersymmetry
to constrain the
form of the multiple D3-brane action on a transverse K\"ahler
manifold.  They compute the first few correction terms in terms of the
curvature tensor $R_{i \bar{j}k \bar{l}}$ of the background.  
They show that the first term, corresponding to our coupling
\begin{equation}
(\partial_{k_1 k_2} h_{ij}) I_h^{ij(k_1 k_2)}
\end{equation}
is uniquely determined by the geodesic length criterion.  The linear
part of the term they find indeed has the same form $\partial^2 h {\rm
Tr}\; X^2 (F^2 + \dot{X}^2)$ as our result for this term.  Their
approach is not able to uniquely determine the higher order
terms, but our results are compatible with the general structure of
the linear parts of the structure they find at higher order.  It is
interesting that they are able to determine the form of some of the
quadratic couplings in their work as well as the linear terms.

It is natural to try to extend the results of this paper to determine
the quadratic and higher order couplings between a system of many
D-branes and the supergravity background fields.  In
\cite{Mark-Wati-3} it was suggested that the coupling to $n$th order
terms in the background could be determined by a $n$-loop calculation
in matrix theory.  The results of Okawa and Yoneya on 3-graviton
scattering in matrix theory \cite{Okawa-Yoneya,Okawa-Yoneya-2} seem to
indicate that this may work at least to quadratic order in general
backgrounds, although there are indications
\cite{deg2,Sethi-Stern-2,Lowe-constraints} that there may be problems
with extending this approach to higher order.  In any case, even the
two-loop calculation would be quite challenging to work out for
completely general background configurations, so it would be nice to
find a simpler approach.  The terms coupling the open string fields of
the D-brane system to any number of bulk supergravity fields can of
course in principle be determined by a perturbative string
calculation.  These calculations are quite complicated, however, even
for the linear coupling terms discussed in this paper.  (For recent
work computing the curvature squared terms in the single D-brane
action see \cite{bbg}).  Furthermore,
the results we have given here extend to arbitrary derivatives in the
background fields and contain a correspondingly arbitrary number of
D-brane fields.  It is difficult to imagine reproducing such results
from perturbative string theory.  Despite these difficulties inherent
in a systematic derivation of the higher order coupling terms, it may
be that the symmetries of the theory and the geodesic length criterion
are sufficient to determine the structure of some of the higher-order
terms.  The results of \cite{dko} on K\"ahler manifolds indicate that
it is indeed possible to learn something about about the higher order
terms using this approach.  In this paper we have found that the
combinatorial structure of the higher moment terms is crucial in
fixing the masses of off-diagonal strings in accord with the geodesic
length criterion.  This indicates that this condition will place
strong constraints on the possible form of the higher order couplings
to the background.  Another approach which might help extend the
results here to higher order is to find the symmetry principle which
corresponds to general coordinate invariance for the multiple D0-brane
system.  Because the coordinates enter the theory as matrices, there
are ordering ambiguities in determining how the operators describing
the D0-brane system transform.  If this symmetry could be understood
in a systematic fashion, it would quite possibly uniquely determine
the higher order couplings of the multiple D-brane system to general
backgrounds.

The approach we have taken here to describing matrix theory and
multiple D0-brane systems in curved backgrounds is to assume that the
action in a weak background can be written as a matrix quantum
mechanics theory with a systematic expansion in powers of the
background field strength.  This approach certainly seems valid for
linear couplings to the background.  It is not clear, however, that
such an approach can be extended to all orders.  In
\cite{dos,Douglas-Ooguri}, in fact, it was argued that even on simple
manifolds like K3 or ALE spaces it may not be possible to describe
DLCQ M-theory using a finite size matrix quantum mechanics theory.  We
are not sure at this point how the approach we have taken here fits
with the results of those papers.  One possibility is that the
perturbation expansion we are considering in powers of the background
field will not converge to a well-defined theory when higher order
terms are taken into account and the background is of the form
considered in \cite{dos,Douglas-Ooguri}.  Indeed, the results of those
papers indicate that the expansion in weak backgrounds may break down
at quadratic order in the curvature of the background\footnote{Thanks
to H.\ Ooguri for correspondence on this point.}.  Another possibility is that
either the restriction to light-front coordinates in M-theory or
the fact that in the Seiberg limit the scale of the metric structures
of interest becomes smaller than the string length may have a subtle
effect on the relationship between the supergravity and open string
descriptions of graviton interactions, leading to some modification in
the conclusions of \cite{dos,Douglas-Ooguri}.  It is clearly a very
important question whether a good low-energy description of D0-branes
can be given which maps to M-theory in the Seiberg limit, but we leave
a further resolution of this issue to further work.

In this paper we have focused on the action for a system of D0-branes
in a weak supergravity background.  It is possible, however, to
T-dualize the action we have given here to get an action for
D$p$-branes of arbitrary dimension in a weak background.  One
particularly interesting case is that of D3-branes.  The T-duals of
the currents $I_x$ we have determined in this paper are linear
combinations of the operators in the ${\cal N} = 4,$ $D = 3 + 1$ super
Yang-Mills theory on the world volume of a system of many D3-branes
which lie in the short representations of the $SU (2, 2 | 4)$
superconformal symmetry group of the theory.  These operators, which
couple linearly to the supergravity background fields, play a crucial
role in the simplest version of Maldacena's AdS/CFT correspondence
\cite{Maldacena-conjecture}.  For the fields associated with the
lowest partial waves of bulk fields in the AdS space, some of the
corresponding operators were found from the Born-Infeld action in
\cite{Klebanov-absorption,gkt,flz,Das-Trivedi}.  All the other lowest
partial wave operators are in principle determined by supersymmetry
and group theory from the weight 2 operator $ {\rm Tr}\; X^{(i}
X^{j)}$, following \cite{krn,Gunaydin-Marcus,gkp-2,Witten-AdS1} and
related work.  Although these operators have played a fundamental role
in understanding the detailed structure of the AdS/CFT correspondence,
only a few of these operators have been described explicitly in terms
of the component fields in the D3-brane world-volume theory.  In
addition, to date no general method for understanding the structure of
the higher partial wave operators (which correspond to the higher
moments of our currents $I_x^{\cdots (\cdots)}$ through T-duality and
which are related through supersymmetry to the higher weight chiral
primary operators ${\rm Tr}\; X^{\{i_1} \cdots X^{i_k\}}$) has
been presented in the literature, although some discussion of
particular operators of this type was given in, for example,
\cite{gkt,dfs2}.  In a separate paper we will discuss more details of
the connection between the results presented here and the operators
which are used in the AdS/CFT correspondence for 3-branes.

A feature which emerges from the results of
\cite{Mark-Wati,Dan-Wati-2,Mark-Wati-3} and the extension of this
work in the present paper is the characteristic form of the higher
moments of the currents $I_x$ which couple to the derivatives of the
background supergravity fields.  In general, we find that the $n$th
moment of the current $I_x$ has  contributions of the form
\begin{equation}
I_x^{(i_1 \cdots i_n)} = {\rm Sym} \; (I_x; X^{i_1}, X^{i_2}, \ldots
X^{i_n}) + I_{x ({\rm fermion})}^{(i_1 \cdots i_n)}
%\label{eq:}
\end{equation}
where the first term on the RHS is a bosonic contribution to the
higher moment given by a symmetrized trace of the polynomial giving
the monopole moment of $I_x$ with a product of $n$ $X$'s, and the
second term contains fermionic contributions to the higher moment.
The fact that all the monopole moments of the currents as well as
their higher moments can be expressed in a symmetrized trace form is
reminiscent of Tseytlin's suggestion \cite{Tseytlin} for using the
symmetrized trace to resolve ordering ambiguities in the nonabelian
Born-Infeld action.  Indeed, the components $T^{--}$ of the 11D stress
tensor are precisely the symmetrized $F^4$ terms appearing at fourth
order in the nonabelian Born-Infeld action proposed by Tseytlin.  This
structure may be helpful in trying to predict the form of higher-order
terms in the action without doing explicit matrix theory or string
theory calculations.  This structure is also helpful in understanding
previous work in which higher partial waves of operators on the
D3-brane play a role.  In particular, in \cite{gkt} the rate of
absorption of higher partial waves of minimally coupled scalars in an
extremal D3-brane background was computed in supergravity and compared
to a D3-brane world-volume calculation.  While the analogous
calculations for $s$-wave absorption can be matched precisely
including numerical coefficients, for partial waves with $l > 1$ a
discrepancy was found in \cite{gkt} between the results of these two
calculations.  The authors suggested that this discrepancy might arise
because the higher partial wave operators were not correctly
normalized.  The results we have given here suggest by T-duality that
they indeed used the correct normalization, but that the operators
they used should have a symmetrized trace with respect to orderings of
the fields.  This additional information seems to help resolve the
discrepancy found in \cite{gkt}\footnote{Thanks to Steve Gubser and Igor
Klebanov for discussions on this point}.  
\junk{For example, the operators of the
form $ O ={\rm Tr}\; F F X X$ appearing as part of the second partial
wave operator ${\rm Tr}\; F^2 (X^i X^j -1/6 \delta^{ij} X^2)$ should be
replaced by the symmetrized operator
\begin{equation}
O ={\rm STr}\;  FF X X =
\frac{2}{3} {\rm Tr}\;  FF X X  +
\frac{1}{3} {\rm Tr}\;  FXF X .
%\label{eq:}
\end{equation}
Since the outgoing particles from the vertices associated with the two
terms on the RHS of this expression can be distinguished at large $N$,
the single term $O$ carrying a factor of 1 should be replaced by two
vertices carrying factors of $(2/3)^2 = 4/9$ and $(1/3)^2 = 1/9$
respectively.  Together, these two vertices seem to cancel the factor
of 9/5 discrepancy found in \cite{gkt} for $l = 2$, although there are
several recalcitrant factors of 2 appearing in the analysis which
complicate this story.  A similar argument gives
combinatorial factors which seem to fix the discrepancy in the case $l = 
3$,
again up to some rogue factors of 2.}
A more detailed
discussion of the resolution of this problem will be described in a
future publication.

\section*{Acknowledgments}

We would like to thank Lorenzo Cornalba,
Steve Gubser, Dan Kabat, Igor Klebanov, Morten Krogh and Ricardo Schiappa for
helpful conversations.  The work of MVR is supported in part by the
Natural Sciences and Engineering Research Council of Canada (NSERC).
The work of WT is supported in part by the A.\ P.\ Sloan Foundation
and in part by the DOE through contract \#DE-FC02-94ER40818.

%%%%%%%%%%%%%%%%%%%%%%

\appendix

\section{Supercurrents from matrix theory}

We reproduce here for convenience the matrix theory forms of the
multipole moments of the 11D supercurrent found in
\cite{Dan-Wati-2,Mark-Wati-3}. The stress tensor $T^{IJ}$, membrane
current $J^{IJK}$ and 5-brane current $M^{IJKLMN}$
have integrated components
\begin{eqnarray}
T^{++} &=& {1 \over R}\str\left(\identity\right)\nonumber\\
T^{+i} &=& {1 \over R}\str\left(\dot{X_i}\right)\nonumber\\
T^{+-} &=& {1 \over R}\str\left({1 \over 2} \dot{X_i} \dot{X_i} + {1 
\over 4} 
F_{ij}  
F_{ij} + {1 \over 2} \theta\gamma^i[X^i,\theta]\right)\nonumber\\
T^{ij} &=& {1 \over R}\str\left( \dot{X_i} \dot{X_j} +  F_{ik}  F_{kj} - 
{1 
\over 4} 
\theta\gamma^i[X_j,\theta] - {1 \over 4} 
\theta\gamma^j[X_i,\theta]\right)\nonumber\\
T^{-i} &=& {1 \over R} \str\left({1 \over 2}\dot{X_i}\dot{X_j}\dot{X_j} 
+ 
{1 
\over 
4} \dot{X_i} F_{jk} F_{jk} + F_{ij} F_{jk} \dot{X_k}\right) \nonumber\\ 
& 
& - 
{1 \over 
4R} 
\str\left(\theta_\alpha 
\dot{X_k}[X_m,\theta_\beta]\right)\{\gamma^k\delta_{im} 
+\gamma^i\delta_{mk} -2\gamma^m\delta_{ki} \}_{\alpha \beta}\nonumber\\ 
& 
& - 
{1 
\over 8R} 
\str\left(\theta_{\alpha} F_{kl}[X_m,\theta_{\beta}]\right)\{ 
\gamma^{[iklm]} 
+ 
2 \gamma^{[lm]} \delta_{ki} + 4\delta_{ki}\delta_{lm} \}_{\alpha 
\beta}\nonumber\\  & & + 
{i \over 8R} \tr(\theta \gamma^{[ki]} \theta \; \theta \gamma^k 
\theta)\nonumber\\
T_f^{--} &=& {1 \over 4R} \str\left(F_{ab}F_{bc}F_{cd}F_{da} - {1 \over 
4}F_{ab} 
F_{ab} F_{cd} F_{cd}  + {\theta} \Gamma^a \Gamma^b \Gamma^c F_{ab} 
F_{cd} 
D_a\theta + {\cal O} ({\theta^4})\right)\nonumber\\
J^{+ij} &=& {1 \over 6R} \str\left(F_{ij}\right) \label{eq:currents}\\
J^{+-i} &=& {1 \over 6R} \str\left( F_{ij} \dot{X_j} - {1 \over 2} 
\theta[X_i,\theta] 
+ {1 
\over 4} \theta \gamma^{[ki]} [X_k, \theta]\right)\nonumber\\
J^{ijk} &=& {1 \over 6R} \str\left( \dot{X_i} F_{jk} +  \dot{X_j} 
F_{ki} 
+
\dot{X_k} F_{ij} - {1 \over 4} \theta 
\gamma^{[ijkl]}[X_l,\theta]\right)\nonumber\\
J^{-ij} &=& {1 \over 6R} \str\left(+\dot{X_i} \dot{X_k} F_{kj} - 
\dot{X_j}\dot{X_k} 
F_{ki} - {1 \over 2} \dot{X_k}\dot{X_k} F_{ij} + {1 \over 4}F_{ij} 
F_{kl} 
F_{kl} 
+ F_{ik} F_{kl} F_{lj}\right)\nonumber\\
& & +{1 \over 24R} \str\left(\theta_\alpha 
\dot{X_k}[X_m,\theta_\beta]\right)\{\gamma^{[kijm]} + 
\gamma^{[jm]} \delta_{ki} - \gamma^{[im]} \delta_{kj} + 2 \delta_{jm} 
\delta_{ki} - 2 \delta_{im} \delta_{kj}\}_{\alpha \beta}\nonumber\\
& & + {1 \over 8} \str\left(\theta_{\alpha} 
F_{kl}[X_m,\theta_{\beta}]\right)\{\gamma^{[jkl]} 
\delta_{mi} - \gamma^{[ikl]} \delta_{mj} + 2 \gamma^{[lij]} \delta_{km} 
+ 
2 
\gamma^l \delta_{jk} \delta_{im} - 2 \gamma^l \delta_{ik} 
\delta_{jm}\nonumber\\
& & \hspace{1in} + 2 \gamma^j \delta_{il} \delta_{km} - 2 \gamma^i 
\delta_{jl} 
\delta_{km}\}_{\alpha \beta}\nonumber\\ & & + {i \over 48R} 
\str\left(\theta 
\gamma^{[kij]} \theta \; \theta 
\gamma^k \theta - \theta \gamma^{[ij]} \theta \; \theta 
\theta\right)\nonumber\\
M^{+-ijkl} &=& {1 \over 12R} \str\left(F_{ij}F_{kl} +F_{ik}F_{lj} + 
F_{il}F_{jk} + 
\theta \gamma^{[jkl}[X^{i]},\theta]\right)\nonumber\\
M^{-ijklm} & = & { 5 \over 4R} \str\left( \dot{X}_{[i}F_{jk}F_{lm]}  
-{1 \over 3}\theta\dot{X}^{[i}\gamma^{jkl}[X^{m]},\theta] - {1 \over 6} 
\theta 
F^{[ij}\gamma^{klm]}\gamma^i [X^i,\theta]\right).\nonumber
\end{eqnarray}
Time derivatives are taken with
respect to Minkowski time $t$.  
All expressions have been written in a gauge with $A_0 = 0$.  Gauge
invariance can be restored by replacing $\dot{X}$ with $D_t X$.
Indices $i, j, \ldots$ run from 1 through 9, while indices $a, b,
\ldots$ run from 0 through 9.
In these expressions we have used the definitions $F_{0i} = \dot{X}^i,
F_{ij} = i[X^i, X^j]$.
We do not know of a matrix form for the transverse 5-brane current
components $M^{+ ijklm},M^{ijklm n}$.

The higher multipole moments of these currents are given by
\begin{eqnarray}
T^{IJ (i_1 \cdots i_k)} & = & \sym (T^{IJ}; X^{i_1}, \ldots, X^{i_k}) + 
T_{\rm fermion}^{IJ(i_1 \cdots i_k)} \nonumber\\
J^{IJK (i_1 \cdots i_k)} & = & \sym (J^{IJK}; X^{i_1}, \ldots, X^{i_k}) 
+ 
J_{\rm fermion}^{IJK(i_1 \cdots i_k)}\\
M^{IJKLMN (i_1 \cdots i_k)} & = & \sym (M^{IJKLMN}; X^{i_1}, \ldots, 
X^{i_k}) + M_{\rm fermion}^{IJKLMN(i_1 \cdots i_k)}\nonumber
\end{eqnarray}
where some simple examples of the two-fermion contribution to the
first moment terms are
\begin{eqnarray*}
T_{\rm fermion}^{+i(j)} &=& {1 \over 8R} \tr(\theta \gamma^{[ij]} 
\theta)\\
T_{\rm fermion}^{+-(i)} &=& {1 \over 16R} \tr(-i\theta F_{kl} 
\gamma^{[kli]} 
\theta + 
2i\theta \dot{X_l} 
\gamma^{[li]} \theta)\\
J_{\rm fermion}^{+ij(k)} &=& {i \over 48R} \tr(\theta \gamma^{[ijk]} 
\theta)\\
J_{\rm fermion}^{+-i(j)} &=& {1 \over 48R} \str\left(i\theta \dot{X_k} 
\gamma^{[kij]} \theta 
+ i \theta 
F_{ik} 
\gamma^{[kj]}\theta\right)\\
M_{\rm fermion}^{+-ijkl(m)} &=& -{i \over 16R} \str\left(\theta 
F^{[jk}\gamma^{il]m}\theta\right)
\end{eqnarray*}  
The remaining two-fermion contributions to the first moments and some
four-fermion terms are also determined by the results in
\cite{Mark-Wati-3}.

There are also fermionic components of the supercurrent which couple
to background fermion fields in the supergravity theory.  We have not
discussed these couplings in this paper, but the matrix theory form of
the currents is determined in \cite{Mark-Wati-3}

There is also a 6-brane current appearing in matrix theory related to
nontrivial 11D background metrics.  The components of this current as
well as its first moments are
\begin{eqnarray}
S^{+ijklmn} & = & \frac{1}{R}  \str\left(F_{[ij} F_{kl} F_{mn]}\right) 
\nonumber\\
S^{+ijklmn(p)} & = & \frac{1}{R}  \str\left(F_{[ij} F_{kl} F_{mn]} X_{p} 
-\theta F_{[kl}F_{mn}\gamma_{pqr]}  \theta\right) 
\label{eq:6-current2}\\
S^{ijklmnp} & = & 
\frac{7}{R}  \str\left(F_{[ij} F_{kl} F_{mn} \dot{X}_{p]}+ 
(\theta^2, \theta^4 \; {\rm terms})\right) 
\nonumber\\
S^{ijklmnp(q)} & = & 
\frac{7}{R}  \str\left(F_{[ij} F_{kl} F_{mn} \dot{X}_{p]} X_{q}  
- \theta \, \dot{X}_{[j} F_{kl} F_{mn} \gamma_{pqr]} \theta +{i \over 2} 
\theta 
\,
 F_{[jk} F_{lm}  
F_{np} \gamma_{qr]} \theta
\right) \nonumber
\end{eqnarray}

\bibliographystyle{plain}
%%%\bibliography{papers}

\end{document}